\documentclass[nofootinbib,reprint,amsmath,amssymb,prd,aps,superscriptaddress]{revtex4-1}
\usepackage{graphicx}
\usepackage{dcolumn}
\usepackage{xcolor,colortbl}
\usepackage{multirow}
\usepackage{comment}
\usepackage{enumitem}
\usepackage{booktabs}
\usepackage[caption=false]{subfig}
\usepackage{bm}

\usepackage{hyperref}
\hypersetup{colorlinks, linkcolor={teal},citecolor={blue},urlcolor={blue}}

\newcommand{\dd}{{\rm d}}

\begin{document}

\title{Quasinormal modes of nonlocal gravity black holes}

\author{Rocco D'Agostino}
\email{rocco.dagostino@inaf.it}
\affiliation{INAF -- Osservatorio Astronomico di Roma, Via Frascati 33, 00078 Monte Porzio Catone, Italy}
\affiliation{INFN -- Sezione di Roma 1, Piazzale Aldo Moro 2, 00185 Roma, Italy}

\author{Vittorio~De~Falco}
\email{v.defalco@ssmeridionale.it}
\affiliation{Ministero dell'Istruzione e del Merito (M.I.M., ex M.I.U.R.), Italy}

\begin{abstract}
We present a comprehensive study of the quasinormal modes of a new class of nonlocal static and spherically symmetric black hole (BH) solutions within the framework of the revised Deser-Woodard theory of gravity. These solutions are constructed as linear perturbations of the Schwarzschild spacetime and are characterized by an inverse power-law behavior of the lapse metric function. We derive the radial profiles of the effective potentials corresponding to scalar, electromagnetic and axial gravitational fluctuations on the BH background. Using the WKB method, complemented by Pad\'e approximants to regularize the trend of the effective potential near its peak, we compute the complex quasinormal mode frequencies associated with each type of perturbation. Our results are examined from both mathematical and physical perspectives, and are substantiated with references to existing literature. In particular, we compare the numerical outcomes with the predictions of the Schwarzschild metric to quantify deviations from the framework of general relativity. When all types of perturbations are combined, the relative deviations of the fundamental modes can reach up to $\sim 12\%$. Finally, we discuss the possibility to place observational bounds in the BH parameter space, based on the predicted sensitivities of future gravitational wave detectors.
\end{abstract}

\maketitle

\section{Introduction}The theoretical challenges posed by general relativity (GR) in reconciling quantum and classical scales continue to question our definitive understanding of the gravitational interaction. Despite the remarkable success of GR in describing the curvature of spacetime induced by massive objects, it remains fundamentally incompatible with a quantum formulation of gravity, standing in contrast to all other fundamental interactions \cite{Carlip:2001wq,Ashtekar:2004eh}, which are consistently formulated within the framework of quantum field theory (QFT). This incompatibility becomes particularly problematic in extreme regimes, such as near singularities and Planck-scale energies, where quantum gravitational effects are expected to become significant.

Although recent advances, such as gravitational wave (GW) \cite{LIGOScientific:2016aoc,LIGOScientific:2017vwq,Abac:2025saz} and black hole (BH) observations \cite{EventHorizonTelescope:2019dse,EventHorizonTelescope:2022wkp}, have reinforced the validity of standard GR in strong-field tests, several fundamental issues remain unresolved across different scales \cite{Berti:2015itd,Barack:2018yly}. Einstein's gravity, in its conventional form, is insufficient to fully explain the dynamics and evolution of the Universe. The most striking evidence for this limitation comes from the requirement of the cosmological constant $\Lambda$ in the standard cosmological paradigm to account for the observed accelerated expansion of the Universe, attributed to dark energy \cite{SupernovaCosmologyProject:1998vns,SupernovaSearchTeam:1998fmf,Peebles:2002gy,Frieman:2008sn,DAgostino:2019wko}. However, the interpretation of $\Lambda$ within QFT leads to the well-known \emph{fine-tuning} problem \cite{Weinberg:1988cp,Padmanabhan:2002ji,DAgostino:2022fcx}, where the theoretical prediction for vacuum energy density exceeds the observed value by several orders of magnitude, making it one of the most deep inconsistencies in contemporary theoretical physics. Moreover, there seems to be also the actual need for additional components at galactic scales, such as dark matter, which can further suggest that GR may be an incomplete description of gravity \cite{Sanders:2002pf,Bertone:2004pz}. 

Attempts to address these challenges have led to the development of modified gravity scenarios \cite{Ferraro:2006jd,Sotiriou:2008rp,Clifton:2011jh,Joyce:2014kja,Koyama:2015vza,Nojiri:2017ncd,Capozziello:2019cav,DAgostino:2019hvh,DAgostino:2022tdk,DAgostino:2025kme}, ranging from higher-order curvature theories to extra-dimensional frameworks. Among these approaches, nonlocal theories of gravity \cite{Arkani-Hamed:2002ukf,Deser:2007jk,Nojiri:2007uq,Hehl:2008eu,Deffayet:2009ca,Maggiore:2013mea,Bombacigno:2024lud} have emerged as a promising avenue to address fundamental inconsistencies of GR without introducing new physical fields or resorting to \emph{ad hoc} corrections. Nonlocal modifications of GR have been motivated by various theoretical considerations, including their potential capability to address singularity problems \cite{Biswas:2011ar,Buoninfante:2018mre}, as they offer a natural explanation for cosmic acceleration \cite{Maggiore:2014sia,Dirian:2014ara,Capozziello:2022rac,Capozziello:2023ccw}, and reconcile gravity with QFT avoiding the traditional pitfalls of local higher-derivative theories \cite{Calcagni:2007ru,Modesto:2011kw,Modesto:2017sdr,Boos:2022biz}. Importantly, they provide novel phenomenological predictions that may be testable in future gravitational and cosmological observations, offering a pathway toward a deeper understanding of gravitational physics beyond Einstein's theory.

One of the most intriguing scenarios within the framework of nonlocal gravity is the model proposed by Deser and Woodard \cite{Deser:2007jk}. This theory avoids the fine-tuning problem by modifying the Einstein equations through the inclusion of a nonlocal function of the Ricci scalar, $f(\Box^{-1}R)$, which alters the behavior of gravity on large scales while remaining consistent with the well-tested predictions of GR at smaller scales. Unlike many modified gravity theories that introduce additional fields or extra dimensions, the Deser-Woodard model remains a purely metric-based modification, making it a more natural extension of GR without introducing new degrees of freedom. However, constructing a viable and theoretically consistent nonlocal function remains a key challenge, as it must reproduce the observed cosmic history while satisfying constraints from solar system tests and large-scale structure formation. To this end, the original model was recently revised \cite{Deser:2019lmm} and applied to several cosmological contexts (see, e.g., \cite{Ding:2019rlp,Chen:2019wlu,Jackson:2021mgw,Jackson:2023faq}).

In \cite{DAgostino:2025wgl}, we further advanced the nonlocal gravity framework of the Deser-Woodard model by reformulating the field equations within a suitable tetrad frame, thus significantly simplifying the underlying differential structure. In this context, we derived static and spherically symmetric BH solutions as linear small corrections to the Schwarzschild metric. 
In a separate recent work \cite{DAgostino:2025sta}, we also presented spherically symmetric and traversable wormhole solutions within the same nonlocal context. Collectively, these developments establish a novel setting for testing nonlocal gravity at astrophysical scales. 

The aim of this paper is to investigate the physical implications of the newly derived BH solutions through the study of their quasinormal modes (QNMs), which serve as a powerful tool for probing the gravitational dynamics in strong-field regimes and identifying possible deviations from GR \cite{Dreyer:2003bv,Lin:2024ubg,Koshelev:2024lyu}. QNMs describe the characteristic oscillations of a perturbed BH, offering crucial insights into its stability and dynamics \cite{Chandrasekhar:1984siy,Berti:2009kk}. 
Mathematically, QNMs are characterized by complex frequencies, with real parts corresponding to oscillation frequencies and the imaginary parts related to the dumping timescales. A particular subclass of these fluctuations, namely the gravitational ones, govern the GW signals emitted by compact objects during merger and accretion processes \cite{Kokkotas:1999bd,Konoplya:2011qq}, providing thus a promising avenue to uncover potential observational signatures of nonlocal gravity.

This work is structured as follows. In Sec.~\ref{sec:BH-nonlocal} we review the new class of BH solutions in the Deser-Woodard nonlocal gravity model. In Sec.~\ref{sec:QNMs}, we set the theoretical framework for analyzing the QNMs. In Sec.~\ref{sec:methodology}, we describe the methodology adopted in our study. 
In Sec.~\ref{sec:results}, we present the numerical results and compare our predictions with those of the Schwarzschild BH, analyzing and discussing their mathematical and physical implications. Finally, in Sec.~\ref{sec:end}, we summarize our findings and outline potential future directions.

Throughout this paper, we assume units $c=\hbar=G=M=1$, with $M$ being the BH mass.

\section{A class of Black hole solutions in nonlocal gravity}
\label{sec:BH-nonlocal}

The nonlocal gravity action pertaining to the Deser-Woodard model is given by \cite{Deser:2019lmm}
\begin{equation} \label{eq:nonlocal-action}
S=\dfrac{1}{16\pi}\int \dd^4 x \sqrt{-g}\, R\left[1+f(Y)\right]\,,
\end{equation}
which is characterized by the distortion function $f(Y)$ that encodes the nonlocal gravitational effects. Here, $R$ is the Ricci scalar curvature and 
$g$ stays for the determinant of the metric tensor, $g_{\mu\nu}$.

The dynamics of model \eqref{eq:nonlocal-action} can be conveniently studied in its localized form through four independent auxiliary scalar fields obeying the following equations of motion:
\begin{subequations} \label{eq:dyn-SF}
\begin{align}
    \Box X&=R\,, \label{eq:X} \\ 
    \Box Y&=g^{\mu\nu}\partial_\mu X\partial_\nu X\,, \label{eq:Y}\\
    \Box U&=-2\nabla_\mu (V\nabla^\mu X)\,, \label{eq:U} \\
    \Box V&=R\frac{\dd f}{\dd Y}\,,\label{eq:V}
\end{align}
\end{subequations}
where $\Box\equiv \nabla_\mu \nabla^\mu$ is the d'Alembert operator such that
\begin{equation} 
    \Box u\equiv\frac{1}{\sqrt{-g}}\partial_\alpha\left[\sqrt{-g}\,\partial^\alpha u\right],
    \label{eq:box-operator}
\end{equation}
with $u$ being a generic scalar field.

The variation of the action \eqref{eq:box-operator} with respect to $g_{\mu\nu}$ gives the field equations in vacuum \cite{Deser:2019lmm}:
\begin{align}
\left(G_{\mu\nu}+g_{\mu\nu}\Box-\nabla_\mu \nabla_\nu\right)W+\mathcal{K}_{(\mu\nu)}-\frac{1}{2}g_{\mu\nu}g^{\alpha\beta}\mathcal{K}_{\alpha\beta} =0\,,
\label{eq:FE}
\end{align}
where $G_{\mu\nu}\equiv R_{\mu\nu}-\frac{1}{2}g_{\mu\nu}R$ is the Einstein tensor, and $W:= 1+U+f(Y)$. Moreover,
\begin{equation}
\mathcal{K}_{\mu\nu}:= \partial_\mu X\partial_\nu U +\partial_\mu Y \partial_\nu V+V\partial_\mu X \partial_\nu X \,,
\label{eq:Kmunu}
\end{equation}
and $\mathcal{K}_{(\mu\nu)}\equiv (\mathcal{K}_{\mu\nu}+\mathcal{K}_{\nu\mu})/2$.

One can then solve Eq.~\eqref{eq:FE} for a static and spherically symmetric metric of the form
\begin{equation}
\dd s^2=-A(r) \dd t^2+\frac{\dd r^2}{B(r)} +r^2(\dd\theta^2+\sin^2\theta\, \dd\varphi^2)\,.
    \label{eq:general-metric}
\end{equation}
In this context, we have recently proved that a class of BH solutions is found for \cite{DAgostino:2025wgl}
\begin{subequations}\label{eq:BH_sol}
\begin{align}
    A(r)&=1-\frac{2}{r}-\frac{\alpha }{r^k}\,, \label{eq:BH_solA}\\
    B(r)&=1-\frac{2}{r}+\frac{\alpha}{3^{k} r^{k+1}(r-3)^2} \left\{3^k r \Big{[}k (r-3) (r-2)\right.\notag\\
    &\left.+4 r-9\Big{]}-3 (r-2) (2 r-3) r^k\right\}\,.
    \label{eq:BH_solB}
\end{align}    
\end{subequations}
Here, the first-order deviations from the Schwarzschild metric are measured by the small perturbative parameter $0<\alpha \ll 1$, while $k>1$ is a real constant \cite{DAgostino:2025wgl}\footnote{Following the approach of \cite{DAgostino:2025wgl}, in this work we express all physical quantities up to first-order in $\alpha$.}. 
It can be shown that the spacetime arising from Eqs.~\eqref{eq:BH_solA} and \eqref{eq:BH_solB} is free from essential singularities (except in $r=0$) beyond the event horizon and fulfills the asymptotic flatness condition (see \cite{DAgostino:2025wgl} for more details). 
In fact, the Ricci scalar reads
\begin{align}
    &R(r)=\frac{3\alpha r^{-k-\frac{5}{2}}}{(r-3)^3} \bigg[3 \left(7 k^2-13 k+4\right) r^{\frac{3}{2}}+k(k-1) r^{\frac{7}{2}}\notag\\
    &+4 k (3-2 k) r^{\frac{5}{2}} -\frac{4r^{k+\frac{3}{2}}}{3^{k-1}}+\frac{2r^{k+\frac{1}{2}}}{3^{k-2}} -18 (k-1)^2 r^{\frac{1}{2}}\bigg],
\end{align}
admitting the finite limit
\begin{equation}
    \lim_{r\to3}R(r)=\frac{\alpha k\left(5+6k-k^2\right)}{3^{k+3}}\,.
\end{equation}

The analytical expressions of the auxiliary scalar fields corresponding to the given BH solution are \cite{DAgostino:2025wgl}
\begin{subequations}
\begin{align}
X(r)&=\alpha  \left[\frac{3^{2-k}-3 r^{1-k}}{3-r}+x_1 \ln \left(\frac{r}{r-2}\right)\right],\label{eq:X_sol}\\
Y(r)&=\alpha  \ln \left(\frac{r}{r-2}\right),
\label{eq:Ysol1}\\
V(r)&=\alpha  \left[\frac{\mathcal{P}_k(r)}{r^{2k-2}}+v_1 \ln \left(\frac{r}{r-2}\right)\right],
    \label{eq:V_sol}\\
U(r)&=0\label{eq:U_sol}\,,
\end{align}    
\end{subequations}
where $x_1$ and $v_1$ are arbitrary real constants, and $\mathcal{P}_k(r)$ is a polynomial of order $2k-3$ in the radial coordinate $r$. Finally, the distortion function reads \cite{DAgostino:2025wgl}
\begin{equation}\label{eq:fY}
f(Y)=\frac{\alpha  \left(e^{Y/\alpha }-1\right)}{e^{Y/\alpha }-3}  \left[3^{1-k}-\left(\frac{2}{e^{Y/\alpha }-1}+2\right)^{1-k}\right].
\end{equation}
It is worth noting that, for $\alpha\to 0$, one recovers the Schwarzschild BH background and, thus, standard GR is obtained in the limit of $f(Y)\to 0$.

Requiring simultaneously $A(r)=B(r)=0$, we can compute the event horizon, whose expression is given by
\begin{equation}
    r_{\rm H}=2+\frac{\alpha}{2^{k-1}}\,.
    \label{eq:rH}
\end{equation}
Moreover, one can show that the photon ring is given as
\begin{align}
r_{\rm ps}&=3 \left[1+\alpha \left(\frac{k+2}{2\times 3^k} \right)\right].
\end{align}

\section{Black hole perturbations}
\label{sec:QNMs}
Perturbations over a BH background spacetime \eqref{eq:general-metric} can be studied using the tortoise coordinate, $r_*$ \cite{del-Corral:2022kbk}:
\begin{equation}\label{eq:tortoise_coordinate}
\frac{\dd r_*}{\dd r}=\sqrt{\frac{g_{rr}}{g_{tt}}}\,.
\end{equation}
Specifically, for the metric \eqref{eq:general-metric},  we have
\begin{align}
\frac{\dd r}{\dd r_*}=\sqrt{A(r)B(r)}\,,
\label{eq:tortoise}
\end{align}
where\footnote{It is worth noting that no singularity occurs at $r=3$, since 
\begin{equation}
\lim_{r\to3}\sqrt{A(r)B(r)}=\frac{1}{12} \left[4- 3^{-k} \left(k^2+k+10\right)\alpha\right].    
\end{equation}
}
\begin{align}
    \sqrt{A(r)B(r)}&= 1-\frac{2}{r}+\frac{\alpha}{2(r-3)^2}\left[\frac{3^{1-k} (2-r) (2 r-3)}{r}\right. \notag \\
    &\left.+\frac{k r^2-5 k r+6 k-r^2+10 r-18}{r^k}\right].
\end{align}

It can be shown that fluctuations of a BH metric  obey the time-independent Schrödinger-like equation \cite{Kim:2008zzj}
\begin{equation}\label{eq:SEgeneral}
        \frac{\dd^2 \Psi^{(i)}(r)}{\dd r_*^2}+\Big[\omega^2-V^{(i)}(r)\Big]\Psi^{(i)}(r)=0\,,
    \end{equation}
where $V(r)$ is a radial potential, and the spin index $i$ labels the different types of perturbations, with $i=0,1,2$ for scalar, electromagnetic, and gravitational perturbations\footnote{In general, gravitational perturbations must be divided into axial ($s=2^+$) and polar ($s=2^-$) types. However, we will consider only the former, thereby simplifying the notation to $s=2$.}, respectively. QNMs are complex frequencies $\omega$, satisfying Eq.~\eqref{eq:SEgeneral}, subject to the boundary conditions
\begin{align}
    \Psi \sim e^{\pm i\omega r_\ast}\,, \quad r_\ast\to \pm \infty\,,
    \label{eq:boundary}
\end{align}
where the minus sign corresponds to a pure ingoing wave at the event horizon, while the plus sign corresponds to a pure outgoing wave at spatial infinity.

\subsection{Scalar perturbations}
Let us start by considering perturbations of the metric \eqref{eq:general-metric} in the form of a massless scalar field, $\Phi(t,r,\theta,\varphi)$. Its equation of motion obeys the Klein-Gordon equation
\begin{equation}
    \Box \Phi\equiv \frac{1}{\sqrt{-g}}\partial_\mu(\sqrt{-g}g^{\mu\nu}\partial_\nu\Phi)=0\,.
    \label{eq:KG}
\end{equation}
In a spherically symmetric spacetime, we can write the solution in the form \cite{Konoplya:2011qq}
\begin{equation}
    \Phi(t,r,\theta,\varphi)=e^{-i\omega t}\sum_{\ell, m}\frac{\Psi^{(0)}_\ell(r)}{r}Y_{\ell m}(\theta,\varphi)\,,
    \label{eq:ansatz}
\end{equation}
where $\omega$ represents the frequency of the QNMs. Here, $\Psi^{(0)}_\ell(r)$ is the radial part of the scalar perturbation, while $Y_{\ell m}$ are the spherical harmonics that describe the angular dependence of the perturbation:
\begin{equation}
    Y_{\ell m}(\theta, \varphi) = (-1)^m \sqrt{\frac{(2\ell + 1)}{4\pi} \frac{(\ell - m)!}{(\ell + m)!}} P_{\ell}^m(\cos \theta) e^{im\varphi},
\end{equation}
with $ P_{\ell}^m$ being the associated Legendre polynomials.

Substituting ansatz \eqref{eq:ansatz} into Eq.~\eqref{eq:KG} and integrating over the angular variables leads to \cite{Konoplya:2011qq}
\begin{equation}
     \frac{\dd^2 \Psi^{(0)}}{\dd r_*^2}+\Big[\omega^2-V^{(0)}\Big]\Psi^{(0)}=0\,,
\end{equation}
where $V^{(0)}$ is the effective potential associated with scalar perturbations:
\begin{equation}
V^{(0)}=A(r)\frac{\ell(\ell+1)}{r^2}+\frac{A(r)B'(r)+A'(r)B(r)}{2r}\,,
\label{eq:POT_scalar}\\
\end{equation}
where the prime denotes the derivative with respect to $r$.

\subsection{Electromagnetic perturbations}

Electromagnetic perturbations of a BH spacetime can be described similarly to scalar perturbations, but they involve vector fields. In this case, the behavior of the electromagnetic field $F_{\mu\nu}$ is governed by the Maxwell equations in curved spacetime:
\begin{equation}
    \nabla_\mu F^{\mu\nu}\equiv \nabla_\mu(\partial^\mu A^\nu-\partial^\nu A^\mu)=0\,,
\end{equation}
where $A^\mu$ is the vector potential and we assume the absence of sources. In particular, one can write 
\begin{equation}
    A_\mu(t,r,\theta,\varphi)=e^{-i\omega t}\sum_{\ell, m}\frac{\Psi^{(1)}_{\mu,\ell}(r)}{r}Y_{\ell m}(\theta,\varphi)\,,
\end{equation}
where $\Psi^{(1)}_{\mu,\ell}(r)$ represents the radial part of the electromagnetic perturbation for each component of $A_\mu$. One can distinguish between two states: polar polarization, affecting the $r$ and $\theta$ components and corresponding to electric-type perturbations; axial polarization, affecting the $\varphi$ component, corresponding to the magnetic-type perturbations. Due to the isospectrality property, each component formally satisfies the equation \cite{Cardoso:2019mqo}
\begin{equation}
     \frac{\dd^2 \Psi^{(1)}}{\dd r_*^2}+\Big[\omega^2-V^{(1)}\Big]\Psi_\ell^{(1)}=0\,,
\end{equation}
where the effective potential reads as
\begin{equation}
    V^{(1)}=A(r)\frac{\ell(\ell+1)}{r^2}\,.
    \label{eq:POT_vector}
\end{equation}

\subsection{Gravitational perturbations}
To analyze gravitational perturbations of the nonlocal field equations, we linearly perturb the metric tensor as
\begin{equation}
    g_{\mu\nu}=\bar{g}_{\mu\nu}+h_{\mu\nu}\,,
\end{equation}
where $\bar{g}_{\mu\nu}$ is the background spacetime, while $|h_{\mu\nu}|\ll 1$.
Then, it is essential to distinguish between axial (odd-parity) and polar (even-parity) sectors, with the former corresponding to rotational perturbations and affecting the momentum flow of the gravitational field, and the latter corresponding to compressional perturbations and affecting the mass distribution \cite{Chandrasekhar:1984siy}. 

It is important to note that the auxiliary fields $\{X, U, Y, V\}$ introduced to localize the nonlocal Lagrangian, as well as $W$, are scalar quantities and hence possess only even-parity perturbations. This means that, under parity transformation $(\theta, \phi) \rightarrow (\pi - \theta, \pi + \phi)$, these fields remain invariant, and their perturbations do not contribute to the odd-parity (axial) part of the field equations. Moreover, since the tensor $\mathcal{K}_{\mu\nu}$ given by Eq.~\eqref{eq:Kmunu} is constructed solely from gradients of these scalar fields, it inherits the even-parity structure. Therefore, all perturbative contributions from $W$ and $\mathcal{K}_{\mu\nu}$ vanish under axial fluctuations at linear order. As a result, the field equations \eqref{eq:FE} reduces in the axial sector to $G_{\mu\nu}^{\text{(axial)}}=0$, identical to GR. This implies that the axial perturbations remain unaltered and carry no imprint of the scalar auxiliary fields at the linear order. All nonlocal effects, including potential deviations from GR, are thus encoded entirely in the polar sector of the perturbative dynamics.

To demonstrate that the axial gravitational potential equals that of GR, we follow the approach of \cite{Chen:2021pxd}. The BH metric \eqref{eq:general-metric} together with Eqs.~\eqref{eq:BH_solA} and \eqref{eq:BH_solB} represents a perturbation of the Schwarzschild spacetime, which we refer to as the background metric. We shall then perturb the nonlocal auxiliary fields as $X=\bar{X}+\delta X$, $Y=\bar{Y}+\delta Y$, $U=\bar{U}+\delta U$, $V=\bar{V}+\delta V$ and $W=\bar{W}+\delta W$, where the bar over a quantity denotes its value at the background.
Computing the Klein-Gordon equations \eqref{eq:dyn-SF} at the background, where the Ricci scalar identically vanishes, we find the following solutions for the auxiliary scalar fields:
\begin{subequations}
\begin{align}
\bar X(r)&=-C_X\ln\left(1-\frac{2}{r}\right),\label{eq:ASC1}\\
\bar V(r)&=C_V\ln\left(1-\frac{2}{r}\right),\label{eq:ASC2}\\
\bar Y(r)&=-\left(C_X\bar X+C_Y\right)\ln\left(1-\frac{2}{r}\right), \label{eq:ASC3}\\ 
\bar U(r)&=\left(2C_X\bar V+C_U\right)\ln\left(1-\frac{2}{r}\right),\label{eq:ASC4}
\end{align}
\end{subequations}
where $C_X, C_Y, C_U, C_V$ are constants of integration.

As highlighted earlier, axial perturbations correspond to odd-parity modes. However, since the auxiliary scalar fields are invariant under coordinate transformations and couple only to the even-parity sector of the metric, they can be excluded from the analysis of axial perturbations. Using this observation and  the definition of $W$, we get
\begin{equation}\label{eq:constraints1}
\bar{W}=\textrm{constant}\,.
\end{equation} 
Additionally, from Eq.~\eqref{eq:Kmunu} we obtain
\begin{equation}\label{eq:constraints2}
C_X^2\bar V+C_X C_V\bar X+C_X C_U+C_V C_Y=0\,,
\end{equation} 
which, in view of the asymptotic flatness condition applied to Eqs.~\eqref{eq:ASC1} and \eqref{eq:ASC2}, can be recast as
\begin{equation}
    C_X C_U+C_V C_Y=0\,.
\end{equation}

To better study the perturbations, we frame the problem in the tetrad formalism. A tetrad field is identified by $e^{\mu}_{\hat \alpha}$, where the tetrad indices are denoted by a hat. 
The tetrads satisfy the following relations:
\begin{align}
e_{\mu}^{\hat \alpha}e^{\mu}_{\hat \beta}=\delta^{\hat \alpha}_{\hat \beta}\,,\quad e_{\mu}^{\hat \alpha}e^{\nu}_{\hat \alpha}=\delta^{\nu}_{\mu}\,,\quad g_{\mu\nu}=\eta_{\hat \alpha\beta}e_{\mu}^{\hat \alpha}e_{\nu}^{\beta}\equiv e_{\hat \alpha\mu}e_{\nu}^{\hat \alpha}\,.
\end{align}
We thus project each quantity defined in the spacetime $g_{\mu\nu}$ onto the tangent space via the tetrad basis. The covariant derivative of a generic rank-two object $H_{\mu\nu}$ in the tetrad frame reads \cite{Chen:2021pxd}
\begin{align}
e^{\lambda}_{\hat \gamma}\nabla_\lambda H_{\mu\nu}e_{\hat \alpha}^{\mu}e_{\hat \beta}^{\nu}=\partial_{\hat\gamma} H_{\hat\alpha\hat\beta}-\eta^{\hat\delta\hat\epsilon}\left(\Lambda_{\hat\delta\hat\alpha\hat\gamma}H_{\hat\epsilon\hat\beta}+\Lambda_{\hat\delta\hat\beta\hat\gamma}H_{\hat\alpha\hat\epsilon}\right)\,,\label{eq:derivative-tetrad}
\end{align}
where $\Lambda_{\hat\gamma\hat\alpha\hat\beta}\equiv\nabla_\mu e_{\hat\beta}^{\mu}e_{\hat\alpha\nu}e_{\hat\gamma}^{\nu}$.
Using Eq.~\eqref{eq:derivative-tetrad} in the field equations \eqref{eq:FE}, we find
\begin{align}
&R_{\hat\alpha\hat\beta}W-\partial_{\hat\alpha}\partial_{\hat\beta} W+\Lambda_{\hat\gamma\hat\beta\hat\alpha}\partial^{\hat\gamma} W+\eta_{\hat\alpha\hat\beta}\Big(\Box W-\frac{R}{2}W \nonumber \\
&-\frac{1}{2}\mathcal{K}^\rho_\rho\Big)+\mathcal{K}_{(\hat\alpha\hat\beta)}=0\,.
\end{align}
Now, we must evaluate the above equation in the background. Therefore, using Eqs. \eqref{eq:constraints1} and \eqref{eq:constraints2} and considering the $(\hat\theta\hat\varphi)$ and $(\hat r\hat\varphi)$ components, we obtain
\begin{equation}
R_{\hat\theta\hat\varphi}=R_{\hat r\hat\varphi}=0\,.    
\end{equation}

Therefore, it is evident that the nonlocal terms do not contribute to the axial perturbations. As a result, the axial oscillations yield the same behavior as in GR\footnote{For a detailed analysis of axial perturbations in Schwarzschild spacetime, see for instance \cite{Chandrasekhar:1984siy}.}. 
Thus, the axial metric perturbations evolve according to the standard Regge-Wheeler equation:
\begin{equation}
     \frac{\dd^2 \Psi^{(2)}}{\dd r_*^2}+\Big[\omega^2-V^{(2)}\Big]\Psi^{(2)}=0\,,
\end{equation}
where the effective potential for the metric \eqref{eq:general-metric} reads
\begin{align}\label{eq:axial-potential}
V^{(2)}=\frac{A(r)}{r^2}&\left[
(\ell+2)(\ell-1)-\frac{rA'(r)B(r)}{2 A(r)} \right.\notag\\ 
& \left.-\frac{r B'(r)}{2} +2B(r)\right].
\end{align}  

We report in Appendix~\ref{app:potentials} the explicit expressions for the potentials of the scalar, electromagnetic, and gravitational perturbations applied to our BH metric \eqref{eq:BH_sol}.
    
\section{Methodology}
\label{sec:methodology}

We denote the QNMs of our BH solution by $\omega_{n\ell s}$, where the triad $(n,\ell,s)$ accounts for the overtones, multipoles, and spin indices, respectively\footnote{Notice the constraints $0\leq n\leq\ell$ and $\ell\geq s$.}.
While $\omega_{n\ell s}$ depends on the parameters $\alpha$ and $k$, we omit this explicit dependence for notational simplicity.
In what follows, we describe the methodology adopted in our analysis.

\subsection{WKB method}
\label{sec:WKB}
To determine the QNM frequencies, we employ a semi-analytical approach based on the Wentzel-Kramers-Brillouin (or better known as WKB) approximation \cite{Schutz:1985km}. This method can be applied to effective potentials that resemble a tunneling barrier and remain constant near the event horizon and at spatial infinity.

Specifically, one can write the asymptotic WKB solution to Eq.~\eqref{eq:SEgeneral} for $r\to \pm \infty$ as
\begin{equation}
    \Psi \sim \exp\left[\frac{1}{\epsilon}\sum_{n=0}^\infty  \epsilon^n S_n(r)\right],
    \label{eq:WKB_asymptotic}
\end{equation}
where the parameter $\epsilon$ defines the order of the WKB approximation, with
\begin{equation}
    S_0(r)=\pm\, i \int^r \sqrt{Q(x)}\,dx
\end{equation}
being the leading order. Here, $Q(x)\equiv \omega^2-V(x)$, and $\pm$ correspond to outgoing $(r\to \infty)$ and ingoing $(r\to-\infty)$ waves, respectively, in accordance to the aforementioned boundary conditions~\eqref{eq:boundary}.

The WKB method then consists of matching these asymptotic solutions with that in the intermediate region, which are constructed in the vicinity of the potential peak are bounded by two classical turning points where we have $Q(r)=0$. Across the turning points, the solutions change from oscillatory to exponential behaviors, and one can approximate the potential via a Taylor expansion around its maximum at $r=r_0$:
\begin{equation}
    Q(r)\approx Q_0+\frac{1}{2}Q_2(r-r_0)^2+\mathcal O(r-r_0)^3\,,
\end{equation}
where $Q_0\equiv Q(r_0)$ and $Q_2\equiv \frac{dQ}{dr_*}\big|_{r=r_0}$.
At the leading order, the WKB quantization condition for QNMs is 
\begin{equation}
\frac{iQ_0}{\sqrt{2Q_2}}=n+\frac{1}{2}\,, \quad n=0,\, 1,\, 2,\hdots 
\label{eq:leading_WKB}
\end{equation}
which yields complex frequencies as roots.

The above result is relatively straightforward for being extended to the $N$-th WKB order through \cite{Iyer:1986nq,Konoplya:2003ii}
\begin{equation}
    \frac{i(\omega^2-V_0)}{\sqrt{-2V_2}}-\sum_{j=2}^N\Lambda_j=n+\frac{1}{2}\,,
    \label{eq:WKB-N}
\end{equation}
where $V_0$ and $V_2$ are the values of the effective potential and its second derivatives with respect to $r_*$ at the peak, respectively. To simplify computations, $r_0$ is computed numerically, thereby avoiding the need to expand $\omega$ to the first order in $\alpha$. This approach introduces only a minimal systematic error, which remains well below the fourth decimal digit and rapidly decreases for smaller values of $\alpha$, making it negligible for practical purposes. The correction terms $\Lambda_j$ depend on the value of the potential and its derivatives up to the $N+3$ order in the maximum, which can be calculated as
\begin{align}
    V_n=\frac{\dd r}{\dd r_*}\frac{\dd V_{n-1}}{\dd r}, \quad n=2,3\hdots
\end{align}
For the purpose of our study, we adopt the third-order WKB method (i.e., $N=3$) as developed in \cite{Iyer:1986np}. 

\subsection{Pad\'e approximants}

Although the effective potentials are regular and well-defined at $r=3$ (see Eqs.~\eqref{eq:limit-scalar}, \eqref{eq:limit-electromagnetic} and \eqref{eq:limit-gravitational}), the numerical evaluation of their higher-order derivatives-- required by the WKB method-- can be affected by instabilities in the vicinity of that point. It is important to stress, however, that this behavior does not originate from any physical or geometrical singularity in the underlying spacetime. These numerical instabilities are common in the WKB analyses and can be effectively mitigated using standard techniques, such as Pad\'e approximants \cite{Baker1996}, which improve the convergence and stability of perturbative calculations without altering the physical content of the potential. This method has been widely adopted in the study of QNMs of BH (see, e.g., \cite{Matyjasek:2017psv,Konoplya:2019hlu,Konoplya:2024lch} and references therein), where it helps to avoid spurious divergencies and to ensure stable QNM values by smoothing the behavior of the involved quantities around the potential peak.

The Pad\'e approximant is defined as the ratio of two polynomials of order $p$ and $q$:
\begin{equation}
    \mathcal{P}_{(p,q)}(x)=\frac{\displaystyle\sum_{i=0}^p a_i x^i}{\displaystyle\sum_{j=0}^q b_j x^j}\,,
\end{equation}
and has proven quite effective in extending the radius of convergence beyond that of more traditional expansions, such as Taylor series\footnote{Pad\'e approximation has also been employed in other gravitational and cosmological contexts to stabilize analytical functions with poles and increase the convergence of physical observables (see e.g.,   \cite{Capozziello:2017ddd,Capozziello:2020ctn,Capozziello:2022wgl} for more details).}.
In our analysis, we approximate the potential using the (2,2) Pad\'e parametrization:
\begin{equation}
    V_{(2,2)}(x)=\frac{a_0+a_1 x+a_2 x^2}{1+b_1x+b_2x^2}\,,
    \label{eq:Pade}
\end{equation}
where $x\equiv r-3$. To determine the unknown coefficients, we compute the Taylor series of the potential at $x=0$\footnote{The $(p,q)$ Pad\'e approximant must be matched to a Taylor series of order $p+q$.}:
\begin{align}
    V(x)=\displaystyle\sum_{k=0}^4 c_k x^k+\mathcal{O}(x^5)\,, \quad c_k\equiv \frac{1}{k!}\frac{\dd^k V}{\dd x^k}\Big|_{x=0}\,.
    \label{eq:Taylor}
\end{align}
Then, by imposing 
\begin{equation}
    V(x)-V_{(2,2)}(x)=\mathcal{O}(x^5)\,,
\end{equation}
and matching terms order by order in $x$, we obtain explicit expressions for  the coefficients in Eq.~\eqref{eq:Pade}:
\begin{align}
    &a_0=c_0\,, \quad a_1=c_1+\frac{c_0 c_1 c_3}{c_2^2}\,, \quad a_2=\frac{c_2^3 +c_1^2 c_3 - c_0 c_2 c_3}{c_2^2}\,,\nonumber \\
    & b_1=\frac{c_1 c_3}{c_2^2}\,, \quad b_2= -\frac{c_3}{c_2}\,.
\end{align}
The same procedure can be applied to approximate the derivatives of the potential near the peak, which are necessary for the evaluation of Eq.~\eqref{eq:WKB-N}.

The degree of accuracy of the Pad\'e method is demonstrated in  Appendix~\ref{app:Pade}, where we compare the radial profiles of the effective potentials and their derivatives with the corresponding Pad\'e approximations. It is evident that the Pad\'e technique successfully overcomes the divergences of the original functions around $r=3$.

\begin{table}
\centering
\setlength{\tabcolsep}{0.8em}
\renewcommand{\arraystretch}{1.2}
\begin{minipage}{0.5\textwidth}
\begin{tabular}{|c|c|c|c|}
\hline
 $\bm{s=0}$ & $\omega_{n\ell s}$ & $\omega_{n\ell s}^\text{Sch}$ \\
\hline
$\ell=0$, $n=0$ & $0.0989 - 0.1091i$ & $0.1046 - 0.1152i$  \\
$\ell=1$, $n=0$ & $0.2867 - 0.0963i$ & $0.2911 - 0.0980i$  \\
$\ell=1$, $n=1$ & $0.2564 - 0.2973i$ & $0.2622 - 0.3074i$  \\
$\ell=2$, $n=0$ & $0.4756 - 0.0954i$ & $0.4832 - 0.0968i$  \\
$\ell=2$, $n=1$ & $0.4566 - 0.2890i$ & $0.4632 - 0.2958i$  \\
$\ell=2$, $n=2$ & $0.4226 - 0.4879i$ & $0.4317 - 0.5034i$  \\
\hline
\hline
$\bm{s=1}$ & $\omega_{n\ell s}$ & $\omega_{n\ell s}^\text{Sch}$ \\
\hline
$\ell=1$, $n=0$ & $0.2423 - 0.0914i$ & $0.2459 - 0.0931i$ \\
$\ell=1$, $n=1$ & $0.2061 - 0.2866i$ & $0.2113 - 0.2958i$ \\
$\ell=2$, $n=0$ & $0.4501 - 0.0937i$ & $0.4571 - 0.0951i$ \\
$\ell=2$, $n=1$ & $0.4296 - 0.2843i$ & $0.4358 - 0.2910i$ \\
$\ell=2$, $n=2$ & $0.3937 - 0.4810i$ & $0.4023 - 0.4959i$ \\
$\ell=3$, $n=0$ & $0.6464 - 0.0943i$ & $0.6567 - 0.0956i$ \\
$\ell=3$, $n=1$ & $0.6322 - 0.2844i$ & $0.6415 - 0.2898i$ \\
$\ell=3$, $n=2$ & $0.6059 - 0.4779i$ & $0.6151 - 0.4901i$ \\
$\ell=3$, $n=3$ & $0.5693 - 0.6749i$ & $0.5814 - 0.6955i$ \\
\hline
\hline
$\bm{s=2}$ & $\omega_{n\ell s}$ &  $\omega_{n\ell s}^\text{Sch}$ \\
\hline 
$\ell=2$, $n=0$  & $0.3671 - 0.0878i$ & $0.3732 - 0.0892i$ \\
$\ell=2$, $n=1$  & $0.3402 - 0.2684i$ & $0.3460 - 0.2749i$ \\
$\ell=2$, $n=2$  & $0.2935 - 0.4573i$ & $0.3029 - 0.4711i$ \\
$\ell=3$, $n=0$  & $0.5896 - 0.0914i$ & $0.5993 - 0.0927i$ \\
$\ell=3$, $n=1$  & $0.5738 - 0.2761i$ & $0.5824 - 0.2814i$ \\
$\ell=3$, $n=2$  & $0.5446 - 0.4650i$ & $0.5532 - 0.4767i$ \\
$\ell=3$, $n=3$  & $0.5044 - 0.6581i$ & $0.5157 - 0.6774i$ \\
$\ell=4$, $n=0$  & $0.7961 - 0.0928i$ & $0.8091 - 0.0942i$ \\
$\ell=4$, $n=1$  & $0.7845 - 0.2795i$ & $0.7965 - 0.2844i$ \\
$\ell=4$, $n=2$  & $0.7624 - 0.4687i$ & $0.7736 - 0.4790i$  \\
$\ell=4$, $n=3$  & $0.7313 - 0.6609i$ & $0.7433 - 0.6783i$ \\
$\ell=4$, $n=4$  & $0.6921 - 0.8557i$ & $0.7072 - 0.8813i$ \\
\hline
\end{tabular}
\end{minipage}
\caption{QNM frequencies for scalar $(s=0)$, electromagnetic $(s=1)$, and gravitational $(s=2)$ perturbations of the nonlocal BH solution with $\alpha=0.1$ and $k=2$. The last columns report the relative reference Schwarzschild values, which are chosen as reported in Table II of \cite{Konoplya:2003ii}.}
\label{tab:frequencies}
\end{table}

\begin{table}
\centering
\setlength{\tabcolsep}{0.7em}
\renewcommand{\arraystretch}{1.2}
\begin{tabular}{|c|c|c|c|c|c|}
\hline
$\bm{s=0}$ & $k=1$ & $k=2$ & $k=3$ & $k=4$ & $k=5$  \\
\hline
$\alpha=0.10$ &
13.34\% & 3.07\% & 0.91\% & 0.29\% & 0.13\%\\
$\alpha=0.05$ & 4.38\% &1.30\%  & 0.43\%  & 0.14\% &0.08\%  \\
$\alpha=0.01$ & 0.71\% & 0.25\% & 0.09\%  & 0.04\% & 0.02\%   \\
\hline
\hline
$\bm{s=1}$ & $k=1$ & $k=2$ & $k=3$ & $k=4$ & $k=5$  \\
\hline
$\alpha=0.10$ & 11.54\% & 2.87\% & 0.86\% & 0.27\% & 0.10\%\\
$\alpha=0.05$ & 4.25\% &1.28\%  & 0.41\%  & 0.14\% &0.05\%  \\
$\alpha=0.01$ & 0.73\% & 0.24\% & 0.08\%  & 0.04\% & 0.01\%   \\
\hline
\hline
$\bm{s=2}$ & $k=1$ & $k=2$ & $k=3$ & $k=4$ & $k=5$  \\
\hline
$\alpha=0.10$ & 11.79\%  & 2.89\% & 0.90\% & 0.32\% & 0.14\% \\
$\alpha=0.05$ & 4.27\% & 1.30\% & 0.43\%  & 0.16\% & 0.07\% \\
$\alpha=0.01$ & 0.73\% & 0.24\% & 0.08\% & 0.03\% & 0.01\%   \\
\hline
\hline
$\bm{\Gamma}$ & $k=1$ & $k=2$ & $k=3$ & $k=4$ & $k=5$  \\
\hline
$\alpha=0.10$ & 11.94\% & 2.75\% & 0.87\% & 0.30\% & 0.20\%\\
$\alpha=0.05$ & 3.91\% & 1.17\% & 0.43\% & 0.14\% &  0.13\%\\
$\alpha=0.01$ & 0.65\% & 0.20\% & 0.07\% & 0.05\% & 0.03\%\\
\hline
\end{tabular}
\caption{Average relative deviations from Schwarzschild QNMs presented in Table~\ref{tab:frequencies}. The last block reflects the numerical values displayed in the bottom-right panel of Fig.~\ref{fig:average_error}.}
\label{tab:var_perc}
\end{table}

\subsection{Relative deviations from Schwarzschild metric}
We quantify the relative deviations from the Schwarzschild QNMs $(\omega^{\rm Sch}_{n\ell s})$ by separately analyzing the real and imaginary parts \cite{Volkel:2019muj,DeSimone:2025sgu}:
\begin{subequations}
\begin{align}
\delta \omega^{\rm (R)}_{n\ell s} &\equiv \bigg|\frac{\text{Re}[ \omega_{n\ell s}-\omega^{\rm Sch}_{n\ell s}]}{\text{Re} [\omega^{\rm Sch}_{n\ell s}]} \bigg|\, \label{rel_error_real},\\
\delta \omega^{\rm (I)}_{n\ell s} &\equiv \bigg|\frac{\text{Im}[  \omega_{n\ell s}-\omega^{\rm Sch}_{n\ell s}]}{\text{Im}[\omega^{\rm Sch}_{n\ell s}]}\bigg|\,. \label{rel_error_imag}
\end{align}
\end{subequations}
Therefore, we measure the total relative deviation as
\begin{equation} \label{eq:relative-error}
    \delta\omega_{n\ell s}\equiv \sqrt{\left[\delta \omega^{\rm (R)}_{n\ell s}\right]^2+\left[\delta\omega_{n\ell s}^{(\rm{I})}\right]^2}\,.
\end{equation}

\begin{figure*}
    \centering
    \hbox{
    \includegraphics[width=2.25in]{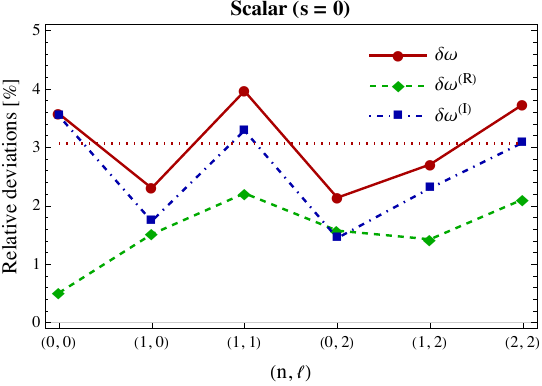}\quad
    \includegraphics[width=2.25in]{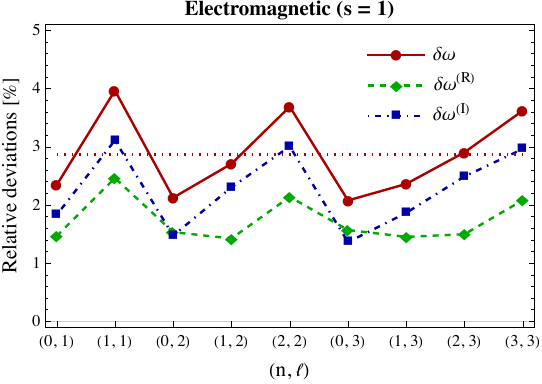}\quad
    \includegraphics[width=2.25in]{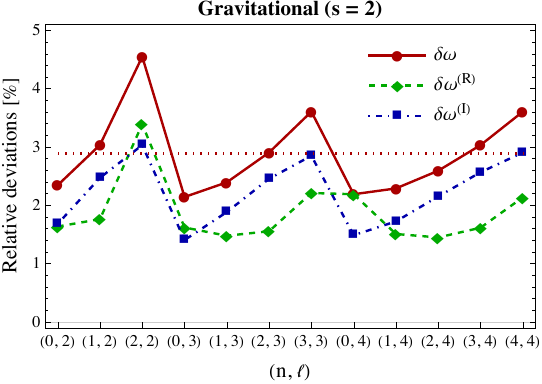}    
    }
    \caption{Relative percentage deviations of the QNM frequencies from their Schwarzschild values for scalar ($s=0$), electromagnetic ($s=1$), and gravitational ($s=2$) perturbations, following the values reported in Table \ref{tab:frequencies} for $\alpha=0.1$ and $k=2$. The total relative deviation (solid red), along with its real (dashed green) and imaginary parts (dot-dashed blue) are computed using Eqs.~\eqref{eq:relative-error}, \eqref{rel_error_real}, and \eqref{rel_error_imag}, respectively. The horizontal dotted lines indicate the average values for each spin sector computed according to Eq.~\eqref{eq:average}: 3.07\%, 2.87\%, 2.89\% for $s=0,1,2$, respectively.}
    \label{fig:all}
\end{figure*}

In the following section, we present the numerical results obtained from our analysis, followed by a discussion of their mathematical and physical implications.

\begin{figure*}
\begin{center}
    \includegraphics[width=3.3in]{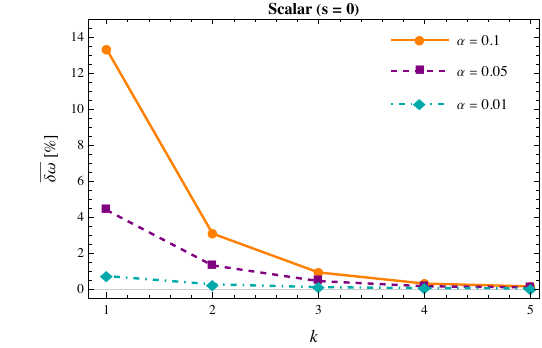} \quad
    \includegraphics[width=3.3in]{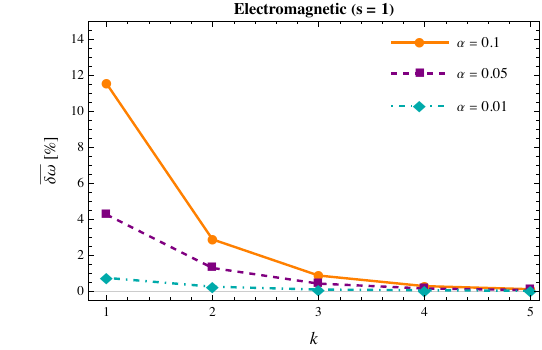}
    \\ \vspace{0.5cm}
    \includegraphics[width=3.3in]{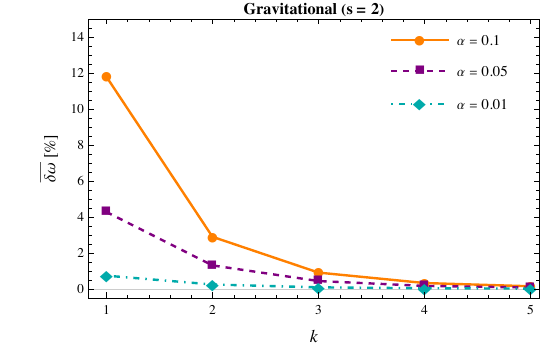} \quad 
    \includegraphics[width=3.3in]{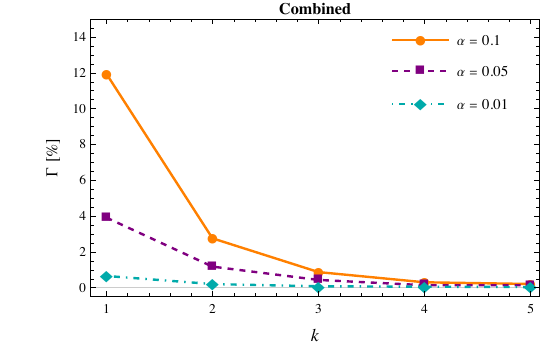}
    \caption{Average percentage deviations from the Schwarzschild predictions of the QNM frequencies for scalar, electromagnetic, and gravitational perturbations of the nonlocal BH solution for different values of the perturbation parameter $\alpha$ (cf. Table~\ref{tab:var_perc}). The bottom-right panel shows the combined relative deviations for the fundamental modes (see Eq.~\eqref{eq:Gamma}).}
    \label{fig:average_error}
\end{center}
\end{figure*}

\section{Results and discussion}
\label{sec:results}
We begin by reporting in Table~\ref{tab:frequencies} the values of the QNM frequencies for scalar, electromagnetic, and gravitational perturbations of the nonlocal BH metric \eqref{eq:BH_sol}, using as demonstrative example $\alpha=0.1$ and $k=2$. For reference, we also include the corresponding entries for the Schwarzschild case \cite{Konoplya:2003ii}. A comparison of the individual frequencies provides a preliminary indication of how close our nonlocal BH solution departs from the GR theory. 

A more detailed analysis, as shown in Fig.~\ref{fig:all}, reveals that the total relative deviation $\delta\omega$ exhibits an oscillatory behavior with respect to $n$ and $\ell$ for all perturbation types. Notably, minima occur when $\ell<n$, while maxima are typically reached at $n=\ell$. This trend can be mainly attributed to the nature of the WKB approximation, which tends to be more accurate when $\ell>n$. Nevertheless, the amplitude of these oscillations can be significantly reduced by increasing $k$ or decreasing $\alpha$.

Also, we note that all three spin sectors show similar qualitative trends in the real and imaginary deviations. For $s=0$, the total deviation peaks at $(n,\ell)=(1,1)$, where both real and imaginary parts are substantial; the real part is generally larger than the imaginary one, indicating that deviations are more sensitive to shifts in the oscillation frequency than damping rate. The case $s=1$ shows smaller fluctuations and more balanced contributions from $\delta\omega^{(\rm R)}$ and  $\delta\omega^{(\rm I)}$; the total average deviation is the lowest among the three cases, reflecting a closer agreement with the Schwarzschild baseline. In the case $s=2$, the highest peak occurs at $(n,\ell)=(2,2)$, where the real part dominates the total deviation; in contrast to the scalar case, modes such as (0,2) and (3,3) reveal a significant contribution from $\delta\omega^{(\rm I)}$, pointing to enhanced deviations in the damping rate.

Therefore, to obtain a clearer measure of the deviations from the Schwarzschild BH across all modes for each spin $s$, we define the average relative deviation as
\begin{equation}
    \overline{\delta\omega}_s\equiv\frac{1}{N_s}\sum_{n,\ell}\delta\omega_{n\ell s}\,,
    \label{eq:average}
\end{equation}
where $N_s$ is the total number of $(n,\ell)$ modes considered in Table~\ref{tab:frequencies} for a given $s$.\footnote{For example, for $s=0$, we have $N_0=6$ and $
\overline{\delta\omega}_0=\frac{1}{6}(\delta\omega_{000}+\delta\omega_{010}+\delta\omega_{110}+\delta\omega_{020}+\delta\omega_{120}+\delta\omega_{220})$.}
We report the resulting outcomes in Table~\ref{tab:var_perc} for different values of the perturbation parameter $\alpha$ and power-law index $k$. As expected, the departures from the Schwarzschild solution progressively decrease with increasing $k$ and decreasing $\alpha$ (see also Fig.~\ref{fig:average_error}). Specifically, for $\alpha=0.1$ and $k=1$, we find maximum deviations of 13.3\%, 11.5\%, and 11.8\% for scalar, electromagnetic, and gravitational perturbations, respectively. These values decrease to $\sim4\%$ for $\alpha=0.05$, and to $\sim 0.7\%$ for $\alpha=0.01$.

Furthermore, to constrain more robustly the deviations from the Schwarzschild metric, we combine the QNMs together for each different type of perturbation. Specifically, we define the following quantity\footnote{Clearly, Eq.~\eqref{eq:Gamma} can be readily generalized to include QNMs with arbitrary values of $n$, $\ell$, and $s$.}
    \begin{equation}
        \Gamma \equiv \frac{\delta \omega_{000}+\delta \omega_{011}+\delta \omega_{022}}{3}\,,
        \label{eq:Gamma}
    \end{equation}
which measures the average relative deviations with respect to the Schwarzschild values among the fundamental (i.e., lowest-overtone) modes of scalar, electromagnetic, and gravitational perturbations.

We report in Table \ref{tab:var_perc} the numerical computations of the $\Gamma$ function in percentage for $\alpha=0.1,0.05,0.01$ and for $k=1,2,3,4,5$. We see that the combined deviations $\Gamma$ can reach up to $11.9\%$ in the case of $k=1$ and $\alpha=0.1$, while they converge to nearly zero for $k\simeq 5$, regardless of the value of the perturbation parameter. This behavior is better visible in Fig.~\ref{fig:average_error}. The order of magnitude of the aforementioned maximum deviations from the Schwarzschild metric is also consistent with the theoretical expectations from the QNM analysis performed in the context of other modified gravity theories \cite{Blazquez-Salcedo:2017txk,Konoplya:2019hml}. In addition, by analyzing both Table~\ref{tab:var_perc} and Fig.~\ref{fig:average_error}, we observe that, for each fixed value of $\alpha$, the $\overline{\delta\omega}$ curves, corresponding to the three different perturbations, take nearly the same values when evaluated at a given $k$. Thus, we can conclude that the degree of departures of our BH solution in terms of QNM frequencies from the Schwarzschild metric remains uniform throughout all the perturbations, as well as in the combined configuration. Ultimately, there is no preferred perturbation to highlight deviations from GR, as all appear to play an equivalent role. This indicates that the combined configuration can provide tighter constraints, as it involves the average of the fundamental modes of each perturbation.

\begin{figure}
    \centering
    \includegraphics[width=3.2in]{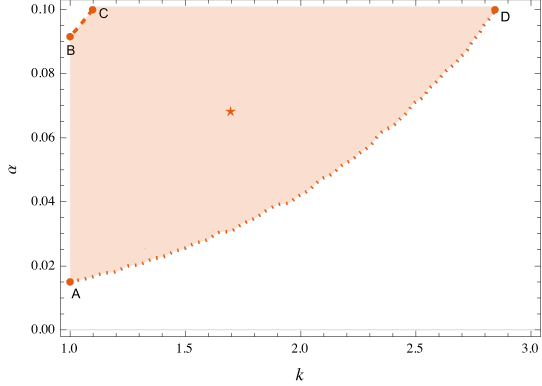}
    \caption{Allowed region in the BH parameter space obtained from the condition $\Gamma\in[1,10]\%$, where the lower and upper bounds are shown as dotted and dashed curves, respectively. The star, located at $(\alpha,k)=(0.07,1.70)$, marks the barycenter of the allowed region, for which $\Gamma=2.41\%$. The coordinates of the vertices are: $A\equiv (1.000,0.015), B\equiv(1.000, 0.092)$, $C\equiv(1.098, 0.100)$ and $D\equiv(2.843, 0.100)$.
    }
    \label{fig:stripe}
\end{figure}

To establish an experimental connection between our theoretical framework and astrophysical data, we identify the allowed region in the $(\alpha,k)$ parameter space by considering the following range of variation for $\Gamma$:
\begin{equation}\label{eq:gamma-range}
    1\% \leq\Gamma(\alpha,k) \leq 10\%\,,
\end{equation}
which roughly corresponds to the expected observational precisions of future GW detectors in measuring deviations from GR predictions during the ringdown phase (see \cite{Bhagwat:2021kwv}, for example). We note that the aforementioned detection amplitude becomes more relevant when one focuses the attention on the coalescence of massive BH binaries (see, e.g., \cite{Berti:2005ys,Berti:2016lat,Barack_2019} for more details). The resulting boundaries are shown in Fig.~\ref{fig:stripe}, yielding $\alpha\gtrsim 0.015$ and $k\in[1,2.84]$. We note that $\alpha$ is not constrained from above. However, the condition $\alpha\ll 1$ must be satisfied to ensure consistency with the underlying assumption of small linear perturbations in our nonlocal BH solution. Therefore, we arbitrarily set the upper bound to $\alpha=0.1$ to better highlight the consequent limits on $k$. 

It is worth emphasizing that the obtained region in the nonlocal BH parameter space is indicative, as it strictly depends on the selected observational window of the involved GW detectors. Our analysis can thus be readapted to different experimental sensibilities, with corresponding changes in the range of Eq.~\eqref{eq:gamma-range}. The chosen $\Gamma$ interval shows that even modest improvements in instrumental precision could further confine or rule out certain sectors of the BH parameter space. Moreover, while the fundamental modes $(n=0)$ are typically dominant in the ringdown signal, higher overtones can also be relevant, especially during the initial phase following the compact object mergers \cite{Giesler:2019uxc,Nee:2023osy}. Therefore, extending the $\Gamma$ definition to include $n>0$ could strengthen the constraints or reveal more complex features that may be negligible or even absent in the case of $n=0$.

\section{Summary and perspectives}
\label{sec:end}
In this work, we investigated perturbations of a class of static and spherically symmetric BH solutions within the framework of the revised Deser-Woodard model of nonlocal gravity. These are formulated as linear fluctuations around the Schwarzschild BH, characterized by a small parameter $\alpha$ and an inverse power-law $g_{tt}$ metric component, parametrized by the index $k$ and a nonlinear function of $k$ for the $g_{rr}$ metric component.
Accordingly, we analyzed the QNM spectrum associated with scalar, electromagnetic, and axial gravitational perturbations over the aforementioned BH spacetime backgrounds. In particular, we showed that the nonlocal effects do not influence the axial perturbations due to parity considerations and, hence, the axial perturbations behave analogously to those in GR. This does not imply that the nonlocal effects are totally absent, but rather that they manifest in the metric as small deviations from GR.

Employing the Regge-Wheeler formalism, we derived the effective potentials corresponding to each type of perturbation. To solve the time-independent Schrodinger-like equation with respect to the tortoise coordinate, we used the third-order WKB method, complemented by the (2,2) Pad\'e approximants to improve the accuracy in the evaluation of the potential near its maximum $r_0$ and the computations of its derivatives always in $r_0$.

To quantify the departures from GR, we calculated the relative differences in the QNM frequencies with respect to the Schwarzschild solution, separating the real and the imaginary parts. For $\alpha=0.1$, the relative deviations from the Schwarzschild BH were found to be $\lesssim (13.3\%,\, 11.5\%,\,11.8\%)$ for scalar, electromagnetic, and gravitational perturbations, respectively. As theoretically expected, the relative deviations progressively reduce with increasing $k$ and decreasing $\alpha$, reaching approximately zero for $k\simeq 5$ across all types of perturbation, independently of the specific value of $\alpha$.

We then introduced the parameter $\Gamma$ to evaluate the combined constraints to unearth the deviations from GR, arising from the fundamental QNMs of the different types of perturbation. In this context, we found $\Gamma< 12\%$.
Furthermore, we placed bounds on the nonlocal BH parameter space using projected sensitivities of future GW detectors. Specifically, requiring $1\%\leq\Gamma\leq 10\%$ yields the constrains $\alpha\gtrsim0.015$ and $k\lesssim 2.84$.

Future works may extend the present analysis along several directions. First, it would be interesting to study polar gravitational perturbations, which may capture significant and additional nonlocal signatures absent in the axial sector \cite{Chen:2021pxd}. Moreover, generalizing the analysis to slowly rotating or Kerr-like nonlocal BHs would considerably enhance the astrophysical relevance of our results, as most astrophysical BHs are expected to be rotating. In this context, BH spectroscopy provides a robust and model-independent method for testing GR in the strong-field regime. Within GR, the complete QNM spectrum of a Kerr BH is uniquely determined by its mass and spin, as stipulated by the no-hair theorem. Any deviation from this spectral structure could signal new physics or modifications to the underlying gravitational theory.

Finally, to bridge theory and observation, it would be important to incorporate Bayesian data analysis techniques that include realistic noise models and detector response functions. Such an approach would enable precise comparisons between nonlocal gravity predictions and GW observations, thereby exploiting the full potential of GW astronomy as a powerful probe of the Kerr geometry and possible deviations from GR. While QNMs serve as direct probes of strong-field dynamics, complementary observables (such as the BH shadow, inspiral waveforms, and gravitational lensing) can further constrain deviations from GR and reduce degeneracies in the parameter space of the nonlocal gravity model under consideration.

\section*{Acknowledgements}
R.D. acknowledges financial support from INFN--Sezione di Roma 1, \textit{esperimento} Euclid. V.D.F. thanks the Gruppo Nazionale di Fisica Matematica of Istituto Nazionale di Alta Matematica for the support.

\bibliography{references}

\begin{thebibliography}{85}%
\makeatletter
\providecommand \@ifxundefined [1]{%
 \@ifx{#1\undefined}
}%
\providecommand \@ifnum [1]{%
 \ifnum #1\expandafter \@firstoftwo
 \else \expandafter \@secondoftwo
 \fi
}%
\providecommand \@ifx [1]{%
 \ifx #1\expandafter \@firstoftwo
 \else \expandafter \@secondoftwo
 \fi
}%
\providecommand \natexlab [1]{#1}%
\providecommand \enquote  [1]{``#1''}%
\providecommand \bibnamefont  [1]{#1}%
\providecommand \bibfnamefont [1]{#1}%
\providecommand \citenamefont [1]{#1}%
\providecommand \href@noop [0]{\@secondoftwo}%
\providecommand \href [0]{\begingroup \@sanitize@url \@href}%
\providecommand \@href[1]{\@@startlink{#1}\@@href}%
\providecommand \@@href[1]{\endgroup#1\@@endlink}%
\providecommand \@sanitize@url [0]{\catcode `\\12\catcode `\$12\catcode `\&12\catcode `\#12\catcode `\^12\catcode `\_12\catcode `\%12\relax}%
\providecommand \@@startlink[1]{}%
\providecommand \@@endlink[0]{}%
\providecommand \url  [0]{\begingroup\@sanitize@url \@url }%
\providecommand \@url [1]{\endgroup\@href {#1}{\urlprefix }}%
\providecommand \urlprefix  [0]{URL }%
\providecommand \Eprint [0]{\href }%
\providecommand \doibase [0]{http://dx.doi.org/}%
\providecommand \selectlanguage [0]{\@gobble}%
\providecommand \bibinfo  [0]{\@secondoftwo}%
\providecommand \bibfield  [0]{\@secondoftwo}%
\providecommand \translation [1]{[#1]}%
\providecommand \BibitemOpen [0]{}%
\providecommand \bibitemStop [0]{}%
\providecommand \bibitemNoStop [0]{.\EOS\space}%
\providecommand \EOS [0]{\spacefactor3000\relax}%
\providecommand \BibitemShut  [1]{\csname bibitem#1\endcsname}%
\let\auto@bib@innerbib\@empty
\bibitem [{\citenamefont {Carlip}(2001)}]{Carlip:2001wq}%
  \BibitemOpen
  \bibfield  {author} {\bibinfo {author} {\bibfnamefont {S.}~\bibnamefont {Carlip}},\ }\href {\doibase 10.1088/0034-4885/64/8/301} {\bibfield  {journal} {\bibinfo  {journal} {Rept. Prog. Phys.}\ }\textbf {\bibinfo {volume} {64}},\ \bibinfo {pages} {885} (\bibinfo {year} {2001})},\ \Eprint {http://arxiv.org/abs/gr-qc/0108040} {arXiv:gr-qc/0108040} \BibitemShut {NoStop}%
\bibitem [{\citenamefont {Ashtekar}\ and\ \citenamefont {Lewandowski}(2004)}]{Ashtekar:2004eh}%
  \BibitemOpen
  \bibfield  {author} {\bibinfo {author} {\bibfnamefont {A.}~\bibnamefont {Ashtekar}}\ and\ \bibinfo {author} {\bibfnamefont {J.}~\bibnamefont {Lewandowski}},\ }\href {\doibase 10.1088/0264-9381/21/15/R01} {\bibfield  {journal} {\bibinfo  {journal} {Class. Quant. Grav.}\ }\textbf {\bibinfo {volume} {21}},\ \bibinfo {pages} {R53} (\bibinfo {year} {2004})},\ \Eprint {http://arxiv.org/abs/gr-qc/0404018} {arXiv:gr-qc/0404018} \BibitemShut {NoStop}%
\bibitem [{\citenamefont {Abbott}\ \emph {et~al.}(2016)\citenamefont {Abbott} \emph {et~al.}}]{LIGOScientific:2016aoc}%
  \BibitemOpen
  \bibfield  {author} {\bibinfo {author} {\bibfnamefont {B.~P.}\ \bibnamefont {Abbott}} \emph {et~al.} (\bibinfo {collaboration} {LIGO Scientific, Virgo}),\ }\href {\doibase 10.1103/PhysRevLett.116.061102} {\bibfield  {journal} {\bibinfo  {journal} {Phys. Rev. Lett.}\ }\textbf {\bibinfo {volume} {116}},\ \bibinfo {pages} {061102} (\bibinfo {year} {2016})},\ \Eprint {http://arxiv.org/abs/1602.03837} {arXiv:1602.03837 [gr-qc]} \BibitemShut {NoStop}%
\bibitem [{\citenamefont {Abbott}\ \emph {et~al.}(2017)\citenamefont {Abbott} \emph {et~al.}}]{LIGOScientific:2017vwq}%
  \BibitemOpen
  \bibfield  {author} {\bibinfo {author} {\bibfnamefont {B.~P.}\ \bibnamefont {Abbott}} \emph {et~al.} (\bibinfo {collaboration} {LIGO Scientific, Virgo}),\ }\href {\doibase 10.1103/PhysRevLett.119.161101} {\bibfield  {journal} {\bibinfo  {journal} {Phys. Rev. Lett.}\ }\textbf {\bibinfo {volume} {119}},\ \bibinfo {pages} {161101} (\bibinfo {year} {2017})},\ \Eprint {http://arxiv.org/abs/1710.05832} {arXiv:1710.05832 [gr-qc]} \BibitemShut {NoStop}%
\bibitem [{\citenamefont {Abac}\ \emph {et~al.}(2025)\citenamefont {Abac} \emph {et~al.}}]{Abac:2025saz}%
  \BibitemOpen
  \bibfield  {author} {\bibinfo {author} {\bibfnamefont {A.}~\bibnamefont {Abac}} \emph {et~al.},\ }\href@noop {} {\  (\bibinfo {year} {2025})},\ \Eprint {http://arxiv.org/abs/2503.12263} {arXiv:2503.12263 [gr-qc]} \BibitemShut {NoStop}%
\bibitem [{\citenamefont {Akiyama}\ \emph {et~al.}(2019)\citenamefont {Akiyama} \emph {et~al.}}]{EventHorizonTelescope:2019dse}%
  \BibitemOpen
  \bibfield  {author} {\bibinfo {author} {\bibfnamefont {K.}~\bibnamefont {Akiyama}} \emph {et~al.} (\bibinfo {collaboration} {Event Horizon Telescope}),\ }\href {\doibase 10.3847/2041-8213/ab0ec7} {\bibfield  {journal} {\bibinfo  {journal} {Astrophys. J. Lett.}\ }\textbf {\bibinfo {volume} {875}},\ \bibinfo {pages} {L1} (\bibinfo {year} {2019})},\ \Eprint {http://arxiv.org/abs/1906.11238} {arXiv:1906.11238 [astro-ph.GA]} \BibitemShut {NoStop}%
\bibitem [{\citenamefont {Akiyama}\ \emph {et~al.}(2022)\citenamefont {Akiyama} \emph {et~al.}}]{EventHorizonTelescope:2022wkp}%
  \BibitemOpen
  \bibfield  {author} {\bibinfo {author} {\bibfnamefont {K.}~\bibnamefont {Akiyama}} \emph {et~al.} (\bibinfo {collaboration} {Event Horizon Telescope}),\ }\href {\doibase 10.3847/2041-8213/ac6674} {\bibfield  {journal} {\bibinfo  {journal} {Astrophys. J. Lett.}\ }\textbf {\bibinfo {volume} {930}},\ \bibinfo {pages} {L12} (\bibinfo {year} {2022})},\ \Eprint {http://arxiv.org/abs/2311.08680} {arXiv:2311.08680 [astro-ph.HE]} \BibitemShut {NoStop}%
\bibitem [{\citenamefont {Berti}\ \emph {et~al.}(2015)\citenamefont {Berti} \emph {et~al.}}]{Berti:2015itd}%
  \BibitemOpen
  \bibfield  {author} {\bibinfo {author} {\bibfnamefont {E.}~\bibnamefont {Berti}} \emph {et~al.},\ }\href {\doibase 10.1088/0264-9381/32/24/243001} {\bibfield  {journal} {\bibinfo  {journal} {Class. Quant. Grav.}\ }\textbf {\bibinfo {volume} {32}},\ \bibinfo {pages} {243001} (\bibinfo {year} {2015})},\ \Eprint {http://arxiv.org/abs/1501.07274} {arXiv:1501.07274 [gr-qc]} \BibitemShut {NoStop}%
\bibitem [{\citenamefont {Barack}\ \emph {et~al.}(2019)\citenamefont {Barack} \emph {et~al.}}]{Barack:2018yly}%
  \BibitemOpen
  \bibfield  {author} {\bibinfo {author} {\bibfnamefont {L.}~\bibnamefont {Barack}} \emph {et~al.},\ }\href {\doibase 10.1088/1361-6382/ab0587} {\bibfield  {journal} {\bibinfo  {journal} {Class. Quant. Grav.}\ }\textbf {\bibinfo {volume} {36}},\ \bibinfo {pages} {143001} (\bibinfo {year} {2019})},\ \Eprint {http://arxiv.org/abs/1806.05195} {arXiv:1806.05195 [gr-qc]} \BibitemShut {NoStop}%
\bibitem [{\citenamefont {Perlmutter}\ \emph {et~al.}(1999)\citenamefont {Perlmutter} \emph {et~al.}}]{SupernovaCosmologyProject:1998vns}%
  \BibitemOpen
  \bibfield  {author} {\bibinfo {author} {\bibfnamefont {S.}~\bibnamefont {Perlmutter}} \emph {et~al.} (\bibinfo {collaboration} {Supernova Cosmology Project}),\ }\href {\doibase 10.1086/307221} {\bibfield  {journal} {\bibinfo  {journal} {Astrophys. J.}\ }\textbf {\bibinfo {volume} {517}},\ \bibinfo {pages} {565} (\bibinfo {year} {1999})},\ \Eprint {http://arxiv.org/abs/astro-ph/9812133} {arXiv:astro-ph/9812133} \BibitemShut {NoStop}%
\bibitem [{\citenamefont {Riess}\ \emph {et~al.}(1998)\citenamefont {Riess} \emph {et~al.}}]{SupernovaSearchTeam:1998fmf}%
  \BibitemOpen
  \bibfield  {author} {\bibinfo {author} {\bibfnamefont {A.~G.}\ \bibnamefont {Riess}} \emph {et~al.} (\bibinfo {collaboration} {Supernova Search Team}),\ }\href {\doibase 10.1086/300499} {\bibfield  {journal} {\bibinfo  {journal} {Astron. J.}\ }\textbf {\bibinfo {volume} {116}},\ \bibinfo {pages} {1009} (\bibinfo {year} {1998})},\ \Eprint {http://arxiv.org/abs/astro-ph/9805201} {arXiv:astro-ph/9805201} \BibitemShut {NoStop}%
\bibitem [{\citenamefont {Peebles}\ and\ \citenamefont {Ratra}(2003)}]{Peebles:2002gy}%
  \BibitemOpen
  \bibfield  {author} {\bibinfo {author} {\bibfnamefont {P.~J.~E.}\ \bibnamefont {Peebles}}\ and\ \bibinfo {author} {\bibfnamefont {B.}~\bibnamefont {Ratra}},\ }\href {\doibase 10.1103/RevModPhys.75.559} {\bibfield  {journal} {\bibinfo  {journal} {Rev. Mod. Phys.}\ }\textbf {\bibinfo {volume} {75}},\ \bibinfo {pages} {559} (\bibinfo {year} {2003})},\ \Eprint {http://arxiv.org/abs/astro-ph/0207347} {arXiv:astro-ph/0207347} \BibitemShut {NoStop}%
\bibitem [{\citenamefont {Frieman}\ \emph {et~al.}(2008)\citenamefont {Frieman}, \citenamefont {Turner},\ and\ \citenamefont {Huterer}}]{Frieman:2008sn}%
  \BibitemOpen
  \bibfield  {author} {\bibinfo {author} {\bibfnamefont {J.}~\bibnamefont {Frieman}}, \bibinfo {author} {\bibfnamefont {M.}~\bibnamefont {Turner}}, \ and\ \bibinfo {author} {\bibfnamefont {D.}~\bibnamefont {Huterer}},\ }\href {\doibase 10.1146/annurev.astro.46.060407.145243} {\bibfield  {journal} {\bibinfo  {journal} {Ann. Rev. Astron. Astrophys.}\ }\textbf {\bibinfo {volume} {46}},\ \bibinfo {pages} {385} (\bibinfo {year} {2008})},\ \Eprint {http://arxiv.org/abs/0803.0982} {arXiv:0803.0982 [astro-ph]} \BibitemShut {NoStop}%
\bibitem [{\citenamefont {D'Agostino}(2019)}]{DAgostino:2019wko}%
  \BibitemOpen
  \bibfield  {author} {\bibinfo {author} {\bibfnamefont {R.}~\bibnamefont {D'Agostino}},\ }\href {\doibase 10.1103/PhysRevD.99.103524} {\bibfield  {journal} {\bibinfo  {journal} {Phys. Rev. D}\ }\textbf {\bibinfo {volume} {99}},\ \bibinfo {pages} {103524} (\bibinfo {year} {2019})},\ \Eprint {http://arxiv.org/abs/1903.03836} {arXiv:1903.03836 [gr-qc]} \BibitemShut {NoStop}%
\bibitem [{\citenamefont {Weinberg}(1989)}]{Weinberg:1988cp}%
  \BibitemOpen
  \bibfield  {author} {\bibinfo {author} {\bibfnamefont {S.}~\bibnamefont {Weinberg}},\ }\href {\doibase 10.1103/RevModPhys.61.1} {\bibfield  {journal} {\bibinfo  {journal} {Rev. Mod. Phys.}\ }\textbf {\bibinfo {volume} {61}},\ \bibinfo {pages} {1} (\bibinfo {year} {1989})}\BibitemShut {NoStop}%
\bibitem [{\citenamefont {Padmanabhan}(2003)}]{Padmanabhan:2002ji}%
  \BibitemOpen
  \bibfield  {author} {\bibinfo {author} {\bibfnamefont {T.}~\bibnamefont {Padmanabhan}},\ }\href {\doibase 10.1016/S0370-1573(03)00120-0} {\bibfield  {journal} {\bibinfo  {journal} {Phys. Rept.}\ }\textbf {\bibinfo {volume} {380}},\ \bibinfo {pages} {235} (\bibinfo {year} {2003})},\ \Eprint {http://arxiv.org/abs/hep-th/0212290} {arXiv:hep-th/0212290} \BibitemShut {NoStop}%
\bibitem [{\citenamefont {D'Agostino}\ \emph {et~al.}(2022)\citenamefont {D'Agostino}, \citenamefont {Luongo},\ and\ \citenamefont {Muccino}}]{DAgostino:2022fcx}%
  \BibitemOpen
  \bibfield  {author} {\bibinfo {author} {\bibfnamefont {R.}~\bibnamefont {D'Agostino}}, \bibinfo {author} {\bibfnamefont {O.}~\bibnamefont {Luongo}}, \ and\ \bibinfo {author} {\bibfnamefont {M.}~\bibnamefont {Muccino}},\ }\href {\doibase 10.1088/1361-6382/ac8af2} {\bibfield  {journal} {\bibinfo  {journal} {Class. Quant. Grav.}\ }\textbf {\bibinfo {volume} {39}},\ \bibinfo {pages} {195014} (\bibinfo {year} {2022})},\ \Eprint {http://arxiv.org/abs/2204.02190} {arXiv:2204.02190 [gr-qc]} \BibitemShut {NoStop}%
\bibitem [{\citenamefont {Sanders}\ and\ \citenamefont {McGaugh}(2002)}]{Sanders:2002pf}%
  \BibitemOpen
  \bibfield  {author} {\bibinfo {author} {\bibfnamefont {R.~H.}\ \bibnamefont {Sanders}}\ and\ \bibinfo {author} {\bibfnamefont {S.~S.}\ \bibnamefont {McGaugh}},\ }\href {\doibase 10.1146/annurev.astro.40.060401.093923} {\bibfield  {journal} {\bibinfo  {journal} {Ann. Rev. Astron. Astrophys.}\ }\textbf {\bibinfo {volume} {40}},\ \bibinfo {pages} {263} (\bibinfo {year} {2002})},\ \Eprint {http://arxiv.org/abs/astro-ph/0204521} {arXiv:astro-ph/0204521} \BibitemShut {NoStop}%
\bibitem [{\citenamefont {Bertone}\ \emph {et~al.}(2005)\citenamefont {Bertone}, \citenamefont {Hooper},\ and\ \citenamefont {Silk}}]{Bertone:2004pz}%
  \BibitemOpen
  \bibfield  {author} {\bibinfo {author} {\bibfnamefont {G.}~\bibnamefont {Bertone}}, \bibinfo {author} {\bibfnamefont {D.}~\bibnamefont {Hooper}}, \ and\ \bibinfo {author} {\bibfnamefont {J.}~\bibnamefont {Silk}},\ }\href {\doibase 10.1016/j.physrep.2004.08.031} {\bibfield  {journal} {\bibinfo  {journal} {Phys. Rept.}\ }\textbf {\bibinfo {volume} {405}},\ \bibinfo {pages} {279} (\bibinfo {year} {2005})},\ \Eprint {http://arxiv.org/abs/hep-ph/0404175} {arXiv:hep-ph/0404175} \BibitemShut {NoStop}%
\bibitem [{\citenamefont {Ferraro}\ and\ \citenamefont {Fiorini}(2007)}]{Ferraro:2006jd}%
  \BibitemOpen
  \bibfield  {author} {\bibinfo {author} {\bibfnamefont {R.}~\bibnamefont {Ferraro}}\ and\ \bibinfo {author} {\bibfnamefont {F.}~\bibnamefont {Fiorini}},\ }\href {\doibase 10.1103/PhysRevD.75.084031} {\bibfield  {journal} {\bibinfo  {journal} {Phys. Rev. D}\ }\textbf {\bibinfo {volume} {75}},\ \bibinfo {pages} {084031} (\bibinfo {year} {2007})},\ \Eprint {http://arxiv.org/abs/gr-qc/0610067} {arXiv:gr-qc/0610067} \BibitemShut {NoStop}%
\bibitem [{\citenamefont {Sotiriou}\ and\ \citenamefont {Faraoni}(2010)}]{Sotiriou:2008rp}%
  \BibitemOpen
  \bibfield  {author} {\bibinfo {author} {\bibfnamefont {T.~P.}\ \bibnamefont {Sotiriou}}\ and\ \bibinfo {author} {\bibfnamefont {V.}~\bibnamefont {Faraoni}},\ }\href {\doibase 10.1103/RevModPhys.82.451} {\bibfield  {journal} {\bibinfo  {journal} {Rev. Mod. Phys.}\ }\textbf {\bibinfo {volume} {82}},\ \bibinfo {pages} {451} (\bibinfo {year} {2010})},\ \Eprint {http://arxiv.org/abs/0805.1726} {arXiv:0805.1726 [gr-qc]} \BibitemShut {NoStop}%
\bibitem [{\citenamefont {Clifton}\ \emph {et~al.}(2012)\citenamefont {Clifton}, \citenamefont {Ferreira}, \citenamefont {Padilla},\ and\ \citenamefont {Skordis}}]{Clifton:2011jh}%
  \BibitemOpen
  \bibfield  {author} {\bibinfo {author} {\bibfnamefont {T.}~\bibnamefont {Clifton}}, \bibinfo {author} {\bibfnamefont {P.~G.}\ \bibnamefont {Ferreira}}, \bibinfo {author} {\bibfnamefont {A.}~\bibnamefont {Padilla}}, \ and\ \bibinfo {author} {\bibfnamefont {C.}~\bibnamefont {Skordis}},\ }\href {\doibase 10.1016/j.physrep.2012.01.001} {\bibfield  {journal} {\bibinfo  {journal} {Phys. Rept.}\ }\textbf {\bibinfo {volume} {513}},\ \bibinfo {pages} {1} (\bibinfo {year} {2012})},\ \Eprint {http://arxiv.org/abs/1106.2476} {arXiv:1106.2476 [astro-ph.CO]} \BibitemShut {NoStop}%
\bibitem [{\citenamefont {Joyce}\ \emph {et~al.}(2015)\citenamefont {Joyce}, \citenamefont {Jain}, \citenamefont {Khoury},\ and\ \citenamefont {Trodden}}]{Joyce:2014kja}%
  \BibitemOpen
  \bibfield  {author} {\bibinfo {author} {\bibfnamefont {A.}~\bibnamefont {Joyce}}, \bibinfo {author} {\bibfnamefont {B.}~\bibnamefont {Jain}}, \bibinfo {author} {\bibfnamefont {J.}~\bibnamefont {Khoury}}, \ and\ \bibinfo {author} {\bibfnamefont {M.}~\bibnamefont {Trodden}},\ }\href {\doibase 10.1016/j.physrep.2014.12.002} {\bibfield  {journal} {\bibinfo  {journal} {Phys. Rept.}\ }\textbf {\bibinfo {volume} {568}},\ \bibinfo {pages} {1} (\bibinfo {year} {2015})},\ \Eprint {http://arxiv.org/abs/1407.0059} {arXiv:1407.0059 [astro-ph.CO]} \BibitemShut {NoStop}%
\bibitem [{\citenamefont {Koyama}(2016)}]{Koyama:2015vza}%
  \BibitemOpen
  \bibfield  {author} {\bibinfo {author} {\bibfnamefont {K.}~\bibnamefont {Koyama}},\ }\href {\doibase 10.1088/0034-4885/79/4/046902} {\bibfield  {journal} {\bibinfo  {journal} {Rept. Prog. Phys.}\ }\textbf {\bibinfo {volume} {79}},\ \bibinfo {pages} {046902} (\bibinfo {year} {2016})},\ \Eprint {http://arxiv.org/abs/1504.04623} {arXiv:1504.04623 [astro-ph.CO]} \BibitemShut {NoStop}%
\bibitem [{\citenamefont {Nojiri}\ \emph {et~al.}(2017)\citenamefont {Nojiri}, \citenamefont {Odintsov},\ and\ \citenamefont {Oikonomou}}]{Nojiri:2017ncd}%
  \BibitemOpen
  \bibfield  {author} {\bibinfo {author} {\bibfnamefont {S.}~\bibnamefont {Nojiri}}, \bibinfo {author} {\bibfnamefont {S.~D.}\ \bibnamefont {Odintsov}}, \ and\ \bibinfo {author} {\bibfnamefont {V.~K.}\ \bibnamefont {Oikonomou}},\ }\href {\doibase 10.1016/j.physrep.2017.06.001} {\bibfield  {journal} {\bibinfo  {journal} {Phys. Rept.}\ }\textbf {\bibinfo {volume} {692}},\ \bibinfo {pages} {1} (\bibinfo {year} {2017})},\ \Eprint {http://arxiv.org/abs/1705.11098} {arXiv:1705.11098 [gr-qc]} \BibitemShut {NoStop}%
\bibitem [{\citenamefont {Capozziello}\ \emph {et~al.}(2019)\citenamefont {Capozziello}, \citenamefont {D'Agostino},\ and\ \citenamefont {Luongo}}]{Capozziello:2019cav}%
  \BibitemOpen
  \bibfield  {author} {\bibinfo {author} {\bibfnamefont {S.}~\bibnamefont {Capozziello}}, \bibinfo {author} {\bibfnamefont {R.}~\bibnamefont {D'Agostino}}, \ and\ \bibinfo {author} {\bibfnamefont {O.}~\bibnamefont {Luongo}},\ }\href {\doibase 10.1142/S0218271819300167} {\bibfield  {journal} {\bibinfo  {journal} {Int. J. Mod. Phys. D}\ }\textbf {\bibinfo {volume} {28}},\ \bibinfo {pages} {1930016} (\bibinfo {year} {2019})},\ \Eprint {http://arxiv.org/abs/1904.01427} {arXiv:1904.01427 [gr-qc]} \BibitemShut {NoStop}%
\bibitem [{\citenamefont {D'Agostino}\ and\ \citenamefont {Nunes}(2019)}]{DAgostino:2019hvh}%
  \BibitemOpen
  \bibfield  {author} {\bibinfo {author} {\bibfnamefont {R.}~\bibnamefont {D'Agostino}}\ and\ \bibinfo {author} {\bibfnamefont {R.~C.}\ \bibnamefont {Nunes}},\ }\href {\doibase 10.1103/PhysRevD.100.044041} {\bibfield  {journal} {\bibinfo  {journal} {Phys. Rev. D}\ }\textbf {\bibinfo {volume} {100}},\ \bibinfo {pages} {044041} (\bibinfo {year} {2019})},\ \Eprint {http://arxiv.org/abs/1907.05516} {arXiv:1907.05516 [gr-qc]} \BibitemShut {NoStop}%
\bibitem [{\citenamefont {D'Agostino}\ and\ \citenamefont {Nunes}(2022)}]{DAgostino:2022tdk}%
  \BibitemOpen
  \bibfield  {author} {\bibinfo {author} {\bibfnamefont {R.}~\bibnamefont {D'Agostino}}\ and\ \bibinfo {author} {\bibfnamefont {R.~C.}\ \bibnamefont {Nunes}},\ }\href {\doibase 10.1103/PhysRevD.106.124053} {\bibfield  {journal} {\bibinfo  {journal} {Phys. Rev. D}\ }\textbf {\bibinfo {volume} {106}},\ \bibinfo {pages} {124053} (\bibinfo {year} {2022})},\ \Eprint {http://arxiv.org/abs/2210.11935} {arXiv:2210.11935 [gr-qc]} \BibitemShut {NoStop}%
\bibitem [{\citenamefont {D'Agostino}\ and\ \citenamefont {Bajardi}(2025)}]{DAgostino:2025kme}%
  \BibitemOpen
  \bibfield  {author} {\bibinfo {author} {\bibfnamefont {R.}~\bibnamefont {D'Agostino}}\ and\ \bibinfo {author} {\bibfnamefont {F.}~\bibnamefont {Bajardi}},\ }\href {\doibase 10.1103/PhysRevD.111.104076} {\bibfield  {journal} {\bibinfo  {journal} {Phys. Rev. D}\ }\textbf {\bibinfo {volume} {111}},\ \bibinfo {pages} {104076} (\bibinfo {year} {2025})},\ \Eprint {http://arxiv.org/abs/2505.21359} {arXiv:2505.21359 [gr-qc]} \BibitemShut {NoStop}%
\bibitem [{\citenamefont {Arkani-Hamed}\ \emph {et~al.}(2002)\citenamefont {Arkani-Hamed}, \citenamefont {Dimopoulos}, \citenamefont {Dvali},\ and\ \citenamefont {Gabadadze}}]{Arkani-Hamed:2002ukf}%
  \BibitemOpen
  \bibfield  {author} {\bibinfo {author} {\bibfnamefont {N.}~\bibnamefont {Arkani-Hamed}}, \bibinfo {author} {\bibfnamefont {S.}~\bibnamefont {Dimopoulos}}, \bibinfo {author} {\bibfnamefont {G.}~\bibnamefont {Dvali}}, \ and\ \bibinfo {author} {\bibfnamefont {G.}~\bibnamefont {Gabadadze}},\ }\href@noop {} {\  (\bibinfo {year} {2002})},\ \Eprint {http://arxiv.org/abs/hep-th/0209227} {arXiv:hep-th/0209227} \BibitemShut {NoStop}%
\bibitem [{\citenamefont {Deser}\ and\ \citenamefont {Woodard}(2007)}]{Deser:2007jk}%
  \BibitemOpen
  \bibfield  {author} {\bibinfo {author} {\bibfnamefont {S.}~\bibnamefont {Deser}}\ and\ \bibinfo {author} {\bibfnamefont {R.~P.}\ \bibnamefont {Woodard}},\ }\href {\doibase 10.1103/PhysRevLett.99.111301} {\bibfield  {journal} {\bibinfo  {journal} {Phys. Rev. Lett.}\ }\textbf {\bibinfo {volume} {99}},\ \bibinfo {pages} {111301} (\bibinfo {year} {2007})},\ \Eprint {http://arxiv.org/abs/0706.2151} {arXiv:0706.2151 [astro-ph]} \BibitemShut {NoStop}%
\bibitem [{\citenamefont {Nojiri}\ and\ \citenamefont {Odintsov}(2008)}]{Nojiri:2007uq}%
  \BibitemOpen
  \bibfield  {author} {\bibinfo {author} {\bibfnamefont {S.}~\bibnamefont {Nojiri}}\ and\ \bibinfo {author} {\bibfnamefont {S.~D.}\ \bibnamefont {Odintsov}},\ }\href {\doibase 10.1016/j.physletb.2007.12.001} {\bibfield  {journal} {\bibinfo  {journal} {Phys. Lett. B}\ }\textbf {\bibinfo {volume} {659}},\ \bibinfo {pages} {821} (\bibinfo {year} {2008})},\ \Eprint {http://arxiv.org/abs/0708.0924} {arXiv:0708.0924 [hep-th]} \BibitemShut {NoStop}%
\bibitem [{\citenamefont {Hehl}\ and\ \citenamefont {Mashhoon}(2009)}]{Hehl:2008eu}%
  \BibitemOpen
  \bibfield  {author} {\bibinfo {author} {\bibfnamefont {F.~W.}\ \bibnamefont {Hehl}}\ and\ \bibinfo {author} {\bibfnamefont {B.}~\bibnamefont {Mashhoon}},\ }\href {\doibase 10.1016/j.physletb.2009.02.033} {\bibfield  {journal} {\bibinfo  {journal} {Phys. Lett. B}\ }\textbf {\bibinfo {volume} {673}},\ \bibinfo {pages} {279} (\bibinfo {year} {2009})},\ \Eprint {http://arxiv.org/abs/0812.1059} {arXiv:0812.1059 [gr-qc]} \BibitemShut {NoStop}%
\bibitem [{\citenamefont {Deffayet}\ and\ \citenamefont {Woodard}(2009)}]{Deffayet:2009ca}%
  \BibitemOpen
  \bibfield  {author} {\bibinfo {author} {\bibfnamefont {C.}~\bibnamefont {Deffayet}}\ and\ \bibinfo {author} {\bibfnamefont {R.~P.}\ \bibnamefont {Woodard}},\ }\href {\doibase 10.1088/1475-7516/2009/08/023} {\bibfield  {journal} {\bibinfo  {journal} {JCAP}\ }\textbf {\bibinfo {volume} {08}},\ \bibinfo {pages} {023} (\bibinfo {year} {2009})},\ \Eprint {http://arxiv.org/abs/0904.0961} {arXiv:0904.0961 [gr-qc]} \BibitemShut {NoStop}%
\bibitem [{\citenamefont {Maggiore}(2014)}]{Maggiore:2013mea}%
  \BibitemOpen
  \bibfield  {author} {\bibinfo {author} {\bibfnamefont {M.}~\bibnamefont {Maggiore}},\ }\href {\doibase 10.1103/PhysRevD.89.043008} {\bibfield  {journal} {\bibinfo  {journal} {Phys. Rev. D}\ }\textbf {\bibinfo {volume} {89}},\ \bibinfo {pages} {043008} (\bibinfo {year} {2014})},\ \Eprint {http://arxiv.org/abs/1307.3898} {arXiv:1307.3898 [hep-th]} \BibitemShut {NoStop}%
\bibitem [{\citenamefont {Bombacigno}\ \emph {et~al.}(2025)\citenamefont {Bombacigno}, \citenamefont {De~Angelis}, \citenamefont {van~de Bruck},\ and\ \citenamefont {Giar{\`e}}}]{Bombacigno:2024lud}%
  \BibitemOpen
  \bibfield  {author} {\bibinfo {author} {\bibfnamefont {F.}~\bibnamefont {Bombacigno}}, \bibinfo {author} {\bibfnamefont {M.}~\bibnamefont {De~Angelis}}, \bibinfo {author} {\bibfnamefont {C.}~\bibnamefont {van~de Bruck}}, \ and\ \bibinfo {author} {\bibfnamefont {W.}~\bibnamefont {Giar{\`e}}},\ }\href {\doibase 10.1088/1475-7516/2025/05/025} {\bibfield  {journal} {\bibinfo  {journal} {JCAP}\ }\textbf {\bibinfo {volume} {05}},\ \bibinfo {pages} {025} (\bibinfo {year} {2025})},\ \Eprint {http://arxiv.org/abs/2412.15064} {arXiv:2412.15064 [hep-th]} \BibitemShut {NoStop}%
\bibitem [{\citenamefont {Biswas}\ \emph {et~al.}(2012)\citenamefont {Biswas}, \citenamefont {Gerwick}, \citenamefont {Koivisto},\ and\ \citenamefont {Mazumdar}}]{Biswas:2011ar}%
  \BibitemOpen
  \bibfield  {author} {\bibinfo {author} {\bibfnamefont {T.}~\bibnamefont {Biswas}}, \bibinfo {author} {\bibfnamefont {E.}~\bibnamefont {Gerwick}}, \bibinfo {author} {\bibfnamefont {T.}~\bibnamefont {Koivisto}}, \ and\ \bibinfo {author} {\bibfnamefont {A.}~\bibnamefont {Mazumdar}},\ }\href {\doibase 10.1103/PhysRevLett.108.031101} {\bibfield  {journal} {\bibinfo  {journal} {Phys. Rev. Lett.}\ }\textbf {\bibinfo {volume} {108}},\ \bibinfo {pages} {031101} (\bibinfo {year} {2012})},\ \Eprint {http://arxiv.org/abs/1110.5249} {arXiv:1110.5249 [gr-qc]} \BibitemShut {NoStop}%
\bibitem [{\citenamefont {Buoninfante}\ \emph {et~al.}(2019)\citenamefont {Buoninfante}, \citenamefont {Lambiase},\ and\ \citenamefont {Mazumdar}}]{Buoninfante:2018mre}%
  \BibitemOpen
  \bibfield  {author} {\bibinfo {author} {\bibfnamefont {L.}~\bibnamefont {Buoninfante}}, \bibinfo {author} {\bibfnamefont {G.}~\bibnamefont {Lambiase}}, \ and\ \bibinfo {author} {\bibfnamefont {A.}~\bibnamefont {Mazumdar}},\ }\href {\doibase 10.1016/j.nuclphysb.2019.114646} {\bibfield  {journal} {\bibinfo  {journal} {Nucl. Phys. B}\ }\textbf {\bibinfo {volume} {944}},\ \bibinfo {pages} {114646} (\bibinfo {year} {2019})},\ \Eprint {http://arxiv.org/abs/1805.03559} {arXiv:1805.03559 [hep-th]} \BibitemShut {NoStop}%
\bibitem [{\citenamefont {Maggiore}\ and\ \citenamefont {Mancarella}(2014)}]{Maggiore:2014sia}%
  \BibitemOpen
  \bibfield  {author} {\bibinfo {author} {\bibfnamefont {M.}~\bibnamefont {Maggiore}}\ and\ \bibinfo {author} {\bibfnamefont {M.}~\bibnamefont {Mancarella}},\ }\href {\doibase 10.1103/PhysRevD.90.023005} {\bibfield  {journal} {\bibinfo  {journal} {Phys. Rev. D}\ }\textbf {\bibinfo {volume} {90}},\ \bibinfo {pages} {023005} (\bibinfo {year} {2014})},\ \Eprint {http://arxiv.org/abs/1402.0448} {arXiv:1402.0448 [hep-th]} \BibitemShut {NoStop}%
\bibitem [{\citenamefont {Dirian}\ \emph {et~al.}(2014)\citenamefont {Dirian}, \citenamefont {Foffa}, \citenamefont {Khosravi}, \citenamefont {Kunz},\ and\ \citenamefont {Maggiore}}]{Dirian:2014ara}%
  \BibitemOpen
  \bibfield  {author} {\bibinfo {author} {\bibfnamefont {Y.}~\bibnamefont {Dirian}}, \bibinfo {author} {\bibfnamefont {S.}~\bibnamefont {Foffa}}, \bibinfo {author} {\bibfnamefont {N.}~\bibnamefont {Khosravi}}, \bibinfo {author} {\bibfnamefont {M.}~\bibnamefont {Kunz}}, \ and\ \bibinfo {author} {\bibfnamefont {M.}~\bibnamefont {Maggiore}},\ }\href {\doibase 10.1088/1475-7516/2014/06/033} {\bibfield  {journal} {\bibinfo  {journal} {JCAP}\ }\textbf {\bibinfo {volume} {06}},\ \bibinfo {pages} {033} (\bibinfo {year} {2014})},\ \Eprint {http://arxiv.org/abs/1403.6068} {arXiv:1403.6068 [astro-ph.CO]} \BibitemShut {NoStop}%
\bibitem [{\citenamefont {Capozziello}\ \emph {et~al.}(2022)\citenamefont {Capozziello}, \citenamefont {D'Agostino},\ and\ \citenamefont {Luongo}}]{Capozziello:2022rac}%
  \BibitemOpen
  \bibfield  {author} {\bibinfo {author} {\bibfnamefont {S.}~\bibnamefont {Capozziello}}, \bibinfo {author} {\bibfnamefont {R.}~\bibnamefont {D'Agostino}}, \ and\ \bibinfo {author} {\bibfnamefont {O.}~\bibnamefont {Luongo}},\ }\href {\doibase 10.1016/j.physletb.2022.137475} {\bibfield  {journal} {\bibinfo  {journal} {Phys. Lett. B}\ }\textbf {\bibinfo {volume} {834}},\ \bibinfo {pages} {137475} (\bibinfo {year} {2022})},\ \Eprint {http://arxiv.org/abs/2207.01276} {arXiv:2207.01276 [gr-qc]} \BibitemShut {NoStop}%
\bibitem [{\citenamefont {Capozziello}\ and\ \citenamefont {D'Agostino}(2023)}]{Capozziello:2023ccw}%
  \BibitemOpen
  \bibfield  {author} {\bibinfo {author} {\bibfnamefont {S.}~\bibnamefont {Capozziello}}\ and\ \bibinfo {author} {\bibfnamefont {R.}~\bibnamefont {D'Agostino}},\ }\href {\doibase 10.1016/j.dark.2023.101346} {\bibfield  {journal} {\bibinfo  {journal} {Phys. Dark Univ.}\ }\textbf {\bibinfo {volume} {42}},\ \bibinfo {pages} {101346} (\bibinfo {year} {2023})},\ \Eprint {http://arxiv.org/abs/2310.03136} {arXiv:2310.03136 [gr-qc]} \BibitemShut {NoStop}%
\bibitem [{\citenamefont {Calcagni}\ \emph {et~al.}(2007)\citenamefont {Calcagni}, \citenamefont {Montobbio},\ and\ \citenamefont {Nardelli}}]{Calcagni:2007ru}%
  \BibitemOpen
  \bibfield  {author} {\bibinfo {author} {\bibfnamefont {G.}~\bibnamefont {Calcagni}}, \bibinfo {author} {\bibfnamefont {M.}~\bibnamefont {Montobbio}}, \ and\ \bibinfo {author} {\bibfnamefont {G.}~\bibnamefont {Nardelli}},\ }\href {\doibase 10.1103/PhysRevD.76.126001} {\bibfield  {journal} {\bibinfo  {journal} {Phys. Rev. D}\ }\textbf {\bibinfo {volume} {76}},\ \bibinfo {pages} {126001} (\bibinfo {year} {2007})},\ \Eprint {http://arxiv.org/abs/0705.3043} {arXiv:0705.3043 [hep-th]} \BibitemShut {NoStop}%
\bibitem [{\citenamefont {Modesto}(2012)}]{Modesto:2011kw}%
  \BibitemOpen
  \bibfield  {author} {\bibinfo {author} {\bibfnamefont {L.}~\bibnamefont {Modesto}},\ }\href {\doibase 10.1103/PhysRevD.86.044005} {\bibfield  {journal} {\bibinfo  {journal} {Phys. Rev. D}\ }\textbf {\bibinfo {volume} {86}},\ \bibinfo {pages} {044005} (\bibinfo {year} {2012})},\ \Eprint {http://arxiv.org/abs/1107.2403} {arXiv:1107.2403 [hep-th]} \BibitemShut {NoStop}%
\bibitem [{\citenamefont {Modesto}\ and\ \citenamefont {Rachwa\l{}}(2017)}]{Modesto:2017sdr}%
  \BibitemOpen
  \bibfield  {author} {\bibinfo {author} {\bibfnamefont {L.}~\bibnamefont {Modesto}}\ and\ \bibinfo {author} {\bibfnamefont {L.}~\bibnamefont {Rachwa\l{}}},\ }\href {\doibase 10.1142/S0218271817300208} {\bibfield  {journal} {\bibinfo  {journal} {Int. J. Mod. Phys. D}\ }\textbf {\bibinfo {volume} {26}},\ \bibinfo {pages} {1730020} (\bibinfo {year} {2017})}\BibitemShut {NoStop}%
\bibitem [{\citenamefont {Boos}\ and\ \citenamefont {Carone}(2023)}]{Boos:2022biz}%
  \BibitemOpen
  \bibfield  {author} {\bibinfo {author} {\bibfnamefont {J.}~\bibnamefont {Boos}}\ and\ \bibinfo {author} {\bibfnamefont {C.~D.}\ \bibnamefont {Carone}},\ }\href {\doibase 10.1007/JHEP06(2023)017} {\bibfield  {journal} {\bibinfo  {journal} {JHEP}\ }\textbf {\bibinfo {volume} {06}},\ \bibinfo {pages} {017} (\bibinfo {year} {2023})},\ \Eprint {http://arxiv.org/abs/2212.00861} {arXiv:2212.00861 [hep-th]} \BibitemShut {NoStop}%
\bibitem [{\citenamefont {Deser}\ and\ \citenamefont {Woodard}(2019)}]{Deser:2019lmm}%
  \BibitemOpen
  \bibfield  {author} {\bibinfo {author} {\bibfnamefont {S.}~\bibnamefont {Deser}}\ and\ \bibinfo {author} {\bibfnamefont {R.~P.}\ \bibnamefont {Woodard}},\ }\href {\doibase 10.1088/1475-7516/2019/06/034} {\bibfield  {journal} {\bibinfo  {journal} {JCAP}\ }\textbf {\bibinfo {volume} {06}},\ \bibinfo {pages} {034} (\bibinfo {year} {2019})},\ \Eprint {http://arxiv.org/abs/1902.08075} {arXiv:1902.08075 [gr-qc]} \BibitemShut {NoStop}%
\bibitem [{\citenamefont {Ding}\ and\ \citenamefont {Deng}(2019)}]{Ding:2019rlp}%
  \BibitemOpen
  \bibfield  {author} {\bibinfo {author} {\bibfnamefont {J.-C.}\ \bibnamefont {Ding}}\ and\ \bibinfo {author} {\bibfnamefont {J.-B.}\ \bibnamefont {Deng}},\ }\href {\doibase 10.1088/1475-7516/2019/12/054} {\bibfield  {journal} {\bibinfo  {journal} {JCAP}\ }\textbf {\bibinfo {volume} {12}},\ \bibinfo {pages} {054} (\bibinfo {year} {2019})},\ \Eprint {http://arxiv.org/abs/1908.11223} {arXiv:1908.11223 [astro-ph.CO]} \BibitemShut {NoStop}%
\bibitem [{\citenamefont {Chen}\ \emph {et~al.}(2019)\citenamefont {Chen}, \citenamefont {Chen},\ and\ \citenamefont {Park}}]{Chen:2019wlu}%
  \BibitemOpen
  \bibfield  {author} {\bibinfo {author} {\bibfnamefont {C.-Y.}\ \bibnamefont {Chen}}, \bibinfo {author} {\bibfnamefont {P.}~\bibnamefont {Chen}}, \ and\ \bibinfo {author} {\bibfnamefont {S.}~\bibnamefont {Park}},\ }\href {\doibase 10.1016/j.physletb.2019.07.024} {\bibfield  {journal} {\bibinfo  {journal} {Phys. Lett. B}\ }\textbf {\bibinfo {volume} {796}},\ \bibinfo {pages} {112} (\bibinfo {year} {2019})},\ \Eprint {http://arxiv.org/abs/1905.04557} {arXiv:1905.04557 [gr-qc]} \BibitemShut {NoStop}%
\bibitem [{\citenamefont {Jackson}\ and\ \citenamefont {Bufalo}(2022)}]{Jackson:2021mgw}%
  \BibitemOpen
  \bibfield  {author} {\bibinfo {author} {\bibfnamefont {D.}~\bibnamefont {Jackson}}\ and\ \bibinfo {author} {\bibfnamefont {R.}~\bibnamefont {Bufalo}},\ }\href {\doibase 10.1088/1475-7516/2022/05/043} {\bibfield  {journal} {\bibinfo  {journal} {JCAP}\ }\textbf {\bibinfo {volume} {05}},\ \bibinfo {pages} {043} (\bibinfo {year} {2022})},\ \Eprint {http://arxiv.org/abs/2110.10008} {arXiv:2110.10008 [gr-qc]} \BibitemShut {NoStop}%
\bibitem [{\citenamefont {Jackson}\ and\ \citenamefont {Bufalo}(2023)}]{Jackson:2023faq}%
  \BibitemOpen
  \bibfield  {author} {\bibinfo {author} {\bibfnamefont {D.}~\bibnamefont {Jackson}}\ and\ \bibinfo {author} {\bibfnamefont {R.}~\bibnamefont {Bufalo}},\ }\href {\doibase 10.1088/1475-7516/2023/05/010} {\bibfield  {journal} {\bibinfo  {journal} {JCAP}\ }\textbf {\bibinfo {volume} {05}},\ \bibinfo {pages} {010} (\bibinfo {year} {2023})},\ \Eprint {http://arxiv.org/abs/2303.17552} {arXiv:2303.17552 [gr-qc]} \BibitemShut {NoStop}%
\bibitem [{\citenamefont {D'Agostino}\ and\ \citenamefont {De~Falco}(2025{\natexlab{a}})}]{DAgostino:2025wgl}%
  \BibitemOpen
  \bibfield  {author} {\bibinfo {author} {\bibfnamefont {R.}~\bibnamefont {D'Agostino}}\ and\ \bibinfo {author} {\bibfnamefont {V.}~\bibnamefont {De~Falco}},\ }\href@noop {} {\  (\bibinfo {year} {2025}{\natexlab{a}})},\ \Eprint {http://arxiv.org/abs/2502.15460} {arXiv:2502.15460 [gr-qc]} \BibitemShut {NoStop}%
\bibitem [{\citenamefont {D'Agostino}\ and\ \citenamefont {De~Falco}(2025{\natexlab{b}})}]{DAgostino:2025sta}%
  \BibitemOpen
  \bibfield  {author} {\bibinfo {author} {\bibfnamefont {R.}~\bibnamefont {D'Agostino}}\ and\ \bibinfo {author} {\bibfnamefont {V.}~\bibnamefont {De~Falco}},\ }\href {\doibase 10.1088/1475-7516/2025/05/069} {\bibfield  {journal} {\bibinfo  {journal} {JCAP}\ }\textbf {\bibinfo {volume} {05}},\ \bibinfo {pages} {069} (\bibinfo {year} {2025}{\natexlab{b}})},\ \Eprint {http://arxiv.org/abs/2501.18007} {arXiv:2501.18007 [gr-qc]} \BibitemShut {NoStop}%
\bibitem [{\citenamefont {Dreyer}\ \emph {et~al.}(2004)\citenamefont {Dreyer}, \citenamefont {Kelly}, \citenamefont {Krishnan}, \citenamefont {Finn}, \citenamefont {Garrison},\ and\ \citenamefont {Lopez-Aleman}}]{Dreyer:2003bv}%
  \BibitemOpen
  \bibfield  {author} {\bibinfo {author} {\bibfnamefont {O.}~\bibnamefont {Dreyer}}, \bibinfo {author} {\bibfnamefont {B.~J.}\ \bibnamefont {Kelly}}, \bibinfo {author} {\bibfnamefont {B.}~\bibnamefont {Krishnan}}, \bibinfo {author} {\bibfnamefont {L.~S.}\ \bibnamefont {Finn}}, \bibinfo {author} {\bibfnamefont {D.}~\bibnamefont {Garrison}}, \ and\ \bibinfo {author} {\bibfnamefont {R.}~\bibnamefont {Lopez-Aleman}},\ }\href {\doibase 10.1088/0264-9381/21/4/003} {\bibfield  {journal} {\bibinfo  {journal} {Class. Quant. Grav.}\ }\textbf {\bibinfo {volume} {21}},\ \bibinfo {pages} {787} (\bibinfo {year} {2004})},\ \Eprint {http://arxiv.org/abs/gr-qc/0309007} {arXiv:gr-qc/0309007} \BibitemShut {NoStop}%
\bibitem [{\citenamefont {Lin}\ \emph {et~al.}(2024)\citenamefont {Lin}, \citenamefont {Bravo-Gaete},\ and\ \citenamefont {Zhang}}]{Lin:2024ubg}%
  \BibitemOpen
  \bibfield  {author} {\bibinfo {author} {\bibfnamefont {J.}~\bibnamefont {Lin}}, \bibinfo {author} {\bibfnamefont {M.}~\bibnamefont {Bravo-Gaete}}, \ and\ \bibinfo {author} {\bibfnamefont {X.}~\bibnamefont {Zhang}},\ }\href {\doibase 10.1103/PhysRevD.109.104039} {\bibfield  {journal} {\bibinfo  {journal} {Phys. Rev. D}\ }\textbf {\bibinfo {volume} {109}},\ \bibinfo {pages} {104039} (\bibinfo {year} {2024})},\ \Eprint {http://arxiv.org/abs/2401.02045} {arXiv:2401.02045 [gr-qc]} \BibitemShut {NoStop}%
\bibitem [{\citenamefont {Koshelev}\ \emph {et~al.}(2024)\citenamefont {Koshelev}, \citenamefont {Li},\ and\ \citenamefont {Tokareva}}]{Koshelev:2024lyu}%
  \BibitemOpen
  \bibfield  {author} {\bibinfo {author} {\bibfnamefont {A.~S.}\ \bibnamefont {Koshelev}}, \bibinfo {author} {\bibfnamefont {C.}~\bibnamefont {Li}}, \ and\ \bibinfo {author} {\bibfnamefont {A.}~\bibnamefont {Tokareva}},\ }\href@noop {} {\  (\bibinfo {year} {2024})},\ \Eprint {http://arxiv.org/abs/2412.02678} {arXiv:2412.02678 [hep-th]} \BibitemShut {NoStop}%
\bibitem [{\citenamefont {Chandrasekhar}(1984)}]{Chandrasekhar:1984siy}%
  \BibitemOpen
  \bibfield  {author} {\bibinfo {author} {\bibfnamefont {S.}~\bibnamefont {Chandrasekhar}},\ }\href {\doibase 10.1007/978-94-009-6469-3_2} {\bibfield  {journal} {\bibinfo  {journal} {Fundam. Theor. Phys.}\ }\textbf {\bibinfo {volume} {9}},\ \bibinfo {pages} {5} (\bibinfo {year} {1984})}\BibitemShut {NoStop}%
\bibitem [{\citenamefont {Berti}\ \emph {et~al.}(2009)\citenamefont {Berti}, \citenamefont {Cardoso},\ and\ \citenamefont {Starinets}}]{Berti:2009kk}%
  \BibitemOpen
  \bibfield  {author} {\bibinfo {author} {\bibfnamefont {E.}~\bibnamefont {Berti}}, \bibinfo {author} {\bibfnamefont {V.}~\bibnamefont {Cardoso}}, \ and\ \bibinfo {author} {\bibfnamefont {A.~O.}\ \bibnamefont {Starinets}},\ }\href {\doibase 10.1088/0264-9381/26/16/163001} {\bibfield  {journal} {\bibinfo  {journal} {Class. Quant. Grav.}\ }\textbf {\bibinfo {volume} {26}},\ \bibinfo {pages} {163001} (\bibinfo {year} {2009})},\ \Eprint {http://arxiv.org/abs/0905.2975} {arXiv:0905.2975 [gr-qc]} \BibitemShut {NoStop}%
\bibitem [{\citenamefont {Kokkotas}\ and\ \citenamefont {Schmidt}(1999)}]{Kokkotas:1999bd}%
  \BibitemOpen
  \bibfield  {author} {\bibinfo {author} {\bibfnamefont {K.~D.}\ \bibnamefont {Kokkotas}}\ and\ \bibinfo {author} {\bibfnamefont {B.~G.}\ \bibnamefont {Schmidt}},\ }\href {\doibase 10.12942/lrr-1999-2} {\bibfield  {journal} {\bibinfo  {journal} {Living Rev. Rel.}\ }\textbf {\bibinfo {volume} {2}},\ \bibinfo {pages} {2} (\bibinfo {year} {1999})},\ \Eprint {http://arxiv.org/abs/gr-qc/9909058} {arXiv:gr-qc/9909058} \BibitemShut {NoStop}%
\bibitem [{\citenamefont {Konoplya}\ and\ \citenamefont {Zhidenko}(2011)}]{Konoplya:2011qq}%
  \BibitemOpen
  \bibfield  {author} {\bibinfo {author} {\bibfnamefont {R.~A.}\ \bibnamefont {Konoplya}}\ and\ \bibinfo {author} {\bibfnamefont {A.}~\bibnamefont {Zhidenko}},\ }\href {\doibase 10.1103/RevModPhys.83.793} {\bibfield  {journal} {\bibinfo  {journal} {Rev. Mod. Phys.}\ }\textbf {\bibinfo {volume} {83}},\ \bibinfo {pages} {793} (\bibinfo {year} {2011})},\ \Eprint {http://arxiv.org/abs/1102.4014} {arXiv:1102.4014 [gr-qc]} \BibitemShut {NoStop}%
\bibitem [{\citenamefont {del Corral}\ and\ \citenamefont {Olmedo}(2022)}]{del-Corral:2022kbk}%
  \BibitemOpen
  \bibfield  {author} {\bibinfo {author} {\bibfnamefont {D.}~\bibnamefont {del Corral}}\ and\ \bibinfo {author} {\bibfnamefont {J.}~\bibnamefont {Olmedo}},\ }\href {\doibase 10.1103/PhysRevD.105.064053} {\bibfield  {journal} {\bibinfo  {journal} {Phys. Rev. D}\ }\textbf {\bibinfo {volume} {105}},\ \bibinfo {pages} {064053} (\bibinfo {year} {2022})},\ \Eprint {http://arxiv.org/abs/2201.09584} {arXiv:2201.09584 [gr-qc]} \BibitemShut {NoStop}%
\bibitem [{\citenamefont {Kim}(2008)}]{Kim:2008zzj}%
  \BibitemOpen
  \bibfield  {author} {\bibinfo {author} {\bibfnamefont {S.-W.}\ \bibnamefont {Kim}},\ }\href {\doibase 10.1143/PTPS.172.21} {\bibfield  {journal} {\bibinfo  {journal} {Prog. Theor. Phys. Suppl.}\ }\textbf {\bibinfo {volume} {172}},\ \bibinfo {pages} {21} (\bibinfo {year} {2008})}\BibitemShut {NoStop}%
\bibitem [{\citenamefont {Cardoso}\ \emph {et~al.}(2019)\citenamefont {Cardoso}, \citenamefont {Kimura}, \citenamefont {Maselli}, \citenamefont {Berti}, \citenamefont {Macedo},\ and\ \citenamefont {McManus}}]{Cardoso:2019mqo}%
  \BibitemOpen
  \bibfield  {author} {\bibinfo {author} {\bibfnamefont {V.}~\bibnamefont {Cardoso}}, \bibinfo {author} {\bibfnamefont {M.}~\bibnamefont {Kimura}}, \bibinfo {author} {\bibfnamefont {A.}~\bibnamefont {Maselli}}, \bibinfo {author} {\bibfnamefont {E.}~\bibnamefont {Berti}}, \bibinfo {author} {\bibfnamefont {C.~F.~B.}\ \bibnamefont {Macedo}}, \ and\ \bibinfo {author} {\bibfnamefont {R.}~\bibnamefont {McManus}},\ }\href {\doibase 10.1103/PhysRevD.99.104077} {\bibfield  {journal} {\bibinfo  {journal} {Phys. Rev. D}\ }\textbf {\bibinfo {volume} {99}},\ \bibinfo {pages} {104077} (\bibinfo {year} {2019})},\ \Eprint {http://arxiv.org/abs/1901.01265} {arXiv:1901.01265 [gr-qc]} \BibitemShut {NoStop}%
\bibitem [{\citenamefont {Chen}\ and\ \citenamefont {Park}(2021)}]{Chen:2021pxd}%
  \BibitemOpen
  \bibfield  {author} {\bibinfo {author} {\bibfnamefont {C.-Y.}\ \bibnamefont {Chen}}\ and\ \bibinfo {author} {\bibfnamefont {S.}~\bibnamefont {Park}},\ }\href {\doibase 10.1103/PhysRevD.103.064029} {\bibfield  {journal} {\bibinfo  {journal} {Phys. Rev. D}\ }\textbf {\bibinfo {volume} {103}},\ \bibinfo {pages} {064029} (\bibinfo {year} {2021})},\ \Eprint {http://arxiv.org/abs/2101.06600} {arXiv:2101.06600 [gr-qc]} \BibitemShut {NoStop}%
\bibitem [{\citenamefont {Schutz}\ and\ \citenamefont {Will}(1985)}]{Schutz:1985km}%
  \BibitemOpen
  \bibfield  {author} {\bibinfo {author} {\bibfnamefont {B.~F.}\ \bibnamefont {Schutz}}\ and\ \bibinfo {author} {\bibfnamefont {C.~M.}\ \bibnamefont {Will}},\ }\href {\doibase 10.1086/184453} {\bibfield  {journal} {\bibinfo  {journal} {Astrophys. J. Lett.}\ }\textbf {\bibinfo {volume} {291}},\ \bibinfo {pages} {L33} (\bibinfo {year} {1985})}\BibitemShut {NoStop}%
\bibitem [{\citenamefont {Iyer}(1987)}]{Iyer:1986nq}%
  \BibitemOpen
  \bibfield  {author} {\bibinfo {author} {\bibfnamefont {S.}~\bibnamefont {Iyer}},\ }\href {\doibase 10.1103/PhysRevD.35.3632} {\bibfield  {journal} {\bibinfo  {journal} {Phys. Rev. D}\ }\textbf {\bibinfo {volume} {35}},\ \bibinfo {pages} {3632} (\bibinfo {year} {1987})}\BibitemShut {NoStop}%
\bibitem [{\citenamefont {Konoplya}(2003)}]{Konoplya:2003ii}%
  \BibitemOpen
  \bibfield  {author} {\bibinfo {author} {\bibfnamefont {R.~A.}\ \bibnamefont {Konoplya}},\ }\href {\doibase 10.1103/PhysRevD.68.024018} {\bibfield  {journal} {\bibinfo  {journal} {Phys. Rev. D}\ }\textbf {\bibinfo {volume} {68}},\ \bibinfo {pages} {024018} (\bibinfo {year} {2003})},\ \Eprint {http://arxiv.org/abs/gr-qc/0303052} {arXiv:gr-qc/0303052} \BibitemShut {NoStop}%
\bibitem [{\citenamefont {Iyer}\ and\ \citenamefont {Will}(1987)}]{Iyer:1986np}%
  \BibitemOpen
  \bibfield  {author} {\bibinfo {author} {\bibfnamefont {S.}~\bibnamefont {Iyer}}\ and\ \bibinfo {author} {\bibfnamefont {C.~M.}\ \bibnamefont {Will}},\ }\href {\doibase 10.1103/PhysRevD.35.3621} {\bibfield  {journal} {\bibinfo  {journal} {Phys. Rev. D}\ }\textbf {\bibinfo {volume} {35}},\ \bibinfo {pages} {3621} (\bibinfo {year} {1987})}\BibitemShut {NoStop}%
\bibitem [{\citenamefont {Baker}\ and\ \citenamefont {Graves-Morris}(1996)}]{Baker1996}%
  \BibitemOpen
  \bibfield  {author} {\bibinfo {author} {\bibfnamefont {G.~A.}\ \bibnamefont {Baker}}\ and\ \bibinfo {author} {\bibfnamefont {P.}~\bibnamefont {Graves-Morris}},\ }\href@noop {} {\emph {\bibinfo {title} {Padé Approximants}}},\ \bibinfo {edition} {2nd}\ ed.,\ Encyclopedia of Mathematics and its Applications\ (\bibinfo  {publisher} {Cambridge University Press},\ \bibinfo {year} {1996})\BibitemShut {NoStop}%
\bibitem [{\citenamefont {Matyjasek}\ and\ \citenamefont {Opala}(2017)}]{Matyjasek:2017psv}%
  \BibitemOpen
  \bibfield  {author} {\bibinfo {author} {\bibfnamefont {J.}~\bibnamefont {Matyjasek}}\ and\ \bibinfo {author} {\bibfnamefont {M.}~\bibnamefont {Opala}},\ }\href {\doibase 10.1103/PhysRevD.96.024011} {\bibfield  {journal} {\bibinfo  {journal} {Phys. Rev. D}\ }\textbf {\bibinfo {volume} {96}},\ \bibinfo {pages} {024011} (\bibinfo {year} {2017})},\ \Eprint {http://arxiv.org/abs/1704.00361} {arXiv:1704.00361 [gr-qc]} \BibitemShut {NoStop}%
\bibitem [{\citenamefont {Konoplya}\ \emph {et~al.}(2019{\natexlab{a}})\citenamefont {Konoplya}, \citenamefont {Zhidenko},\ and\ \citenamefont {Zinhailo}}]{Konoplya:2019hlu}%
  \BibitemOpen
  \bibfield  {author} {\bibinfo {author} {\bibfnamefont {R.~A.}\ \bibnamefont {Konoplya}}, \bibinfo {author} {\bibfnamefont {A.}~\bibnamefont {Zhidenko}}, \ and\ \bibinfo {author} {\bibfnamefont {A.~F.}\ \bibnamefont {Zinhailo}},\ }\href {\doibase 10.1088/1361-6382/ab2e25} {\bibfield  {journal} {\bibinfo  {journal} {Class. Quant. Grav.}\ }\textbf {\bibinfo {volume} {36}},\ \bibinfo {pages} {155002} (\bibinfo {year} {2019}{\natexlab{a}})},\ \Eprint {http://arxiv.org/abs/1904.10333} {arXiv:1904.10333 [gr-qc]} \BibitemShut {NoStop}%
\bibitem [{\citenamefont {Konoplya}\ and\ \citenamefont {Stashko}(2024)}]{Konoplya:2024lch}%
  \BibitemOpen
  \bibfield  {author} {\bibinfo {author} {\bibfnamefont {R.~A.}\ \bibnamefont {Konoplya}}\ and\ \bibinfo {author} {\bibfnamefont {O.~S.}\ \bibnamefont {Stashko}},\ }\href@noop {} {\  (\bibinfo {year} {2024})},\ \Eprint {http://arxiv.org/abs/2408.02578} {arXiv:2408.02578 [gr-qc]} \BibitemShut {NoStop}%
\bibitem [{\citenamefont {Capozziello}\ \emph {et~al.}(2018)\citenamefont {Capozziello}, \citenamefont {D'Agostino},\ and\ \citenamefont {Luongo}}]{Capozziello:2017ddd}%
  \BibitemOpen
  \bibfield  {author} {\bibinfo {author} {\bibfnamefont {S.}~\bibnamefont {Capozziello}}, \bibinfo {author} {\bibfnamefont {R.}~\bibnamefont {D'Agostino}}, \ and\ \bibinfo {author} {\bibfnamefont {O.}~\bibnamefont {Luongo}},\ }\href {\doibase 10.1088/1475-7516/2018/05/008} {\bibfield  {journal} {\bibinfo  {journal} {JCAP}\ }\textbf {\bibinfo {volume} {05}},\ \bibinfo {pages} {008} (\bibinfo {year} {2018})},\ \Eprint {http://arxiv.org/abs/1709.08407} {arXiv:1709.08407 [gr-qc]} \BibitemShut {NoStop}%
\bibitem [{\citenamefont {Capozziello}\ \emph {et~al.}(2020)\citenamefont {Capozziello}, \citenamefont {D'Agostino},\ and\ \citenamefont {Luongo}}]{Capozziello:2020ctn}%
  \BibitemOpen
  \bibfield  {author} {\bibinfo {author} {\bibfnamefont {S.}~\bibnamefont {Capozziello}}, \bibinfo {author} {\bibfnamefont {R.}~\bibnamefont {D'Agostino}}, \ and\ \bibinfo {author} {\bibfnamefont {O.}~\bibnamefont {Luongo}},\ }\href {\doibase 10.1093/mnras/staa871} {\bibfield  {journal} {\bibinfo  {journal} {Mon. Not. Roy. Astron. Soc.}\ }\textbf {\bibinfo {volume} {494}},\ \bibinfo {pages} {2576} (\bibinfo {year} {2020})},\ \Eprint {http://arxiv.org/abs/2003.09341} {arXiv:2003.09341 [astro-ph.CO]} \BibitemShut {NoStop}%
\bibitem [{\citenamefont {Capozziello}\ and\ \citenamefont {D'Agostino}(2022)}]{Capozziello:2022wgl}%
  \BibitemOpen
  \bibfield  {author} {\bibinfo {author} {\bibfnamefont {S.}~\bibnamefont {Capozziello}}\ and\ \bibinfo {author} {\bibfnamefont {R.}~\bibnamefont {D'Agostino}},\ }\href {\doibase 10.1016/j.physletb.2022.137229} {\bibfield  {journal} {\bibinfo  {journal} {Phys. Lett. B}\ }\textbf {\bibinfo {volume} {832}},\ \bibinfo {pages} {137229} (\bibinfo {year} {2022})},\ \Eprint {http://arxiv.org/abs/2204.01015} {arXiv:2204.01015 [gr-qc]} \BibitemShut {NoStop}%
\bibitem [{\citenamefont {V\"olkel}\ and\ \citenamefont {Kokkotas}(2019)}]{Volkel:2019muj}%
  \BibitemOpen
  \bibfield  {author} {\bibinfo {author} {\bibfnamefont {S.~H.}\ \bibnamefont {V\"olkel}}\ and\ \bibinfo {author} {\bibfnamefont {K.~D.}\ \bibnamefont {Kokkotas}},\ }\href {\doibase 10.1103/PhysRevD.100.044026} {\bibfield  {journal} {\bibinfo  {journal} {Phys. Rev. D}\ }\textbf {\bibinfo {volume} {100}},\ \bibinfo {pages} {044026} (\bibinfo {year} {2019})},\ \Eprint {http://arxiv.org/abs/1908.00252} {arXiv:1908.00252 [gr-qc]} \BibitemShut {NoStop}%
\bibitem [{\citenamefont {De~Simone}\ \emph {et~al.}(2025)\citenamefont {De~Simone}, \citenamefont {De~Falco},\ and\ \citenamefont {Capozziello}}]{DeSimone:2025sgu}%
  \BibitemOpen
  \bibfield  {author} {\bibinfo {author} {\bibfnamefont {C.}~\bibnamefont {De~Simone}}, \bibinfo {author} {\bibfnamefont {V.}~\bibnamefont {De~Falco}}, \ and\ \bibinfo {author} {\bibfnamefont {S.}~\bibnamefont {Capozziello}},\ }\href {\doibase 10.1103/PhysRevD.111.064021} {\bibfield  {journal} {\bibinfo  {journal} {Phys. Rev. D}\ }\textbf {\bibinfo {volume} {111}},\ \bibinfo {pages} {064021} (\bibinfo {year} {2025})},\ \Eprint {http://arxiv.org/abs/2502.12646} {arXiv:2502.12646 [gr-qc]} \BibitemShut {NoStop}%
\bibitem [{\citenamefont {Bl\'azquez-Salcedo}\ \emph {et~al.}(2017)\citenamefont {Bl\'azquez-Salcedo}, \citenamefont {Khoo},\ and\ \citenamefont {Kunz}}]{Blazquez-Salcedo:2017txk}%
  \BibitemOpen
  \bibfield  {author} {\bibinfo {author} {\bibfnamefont {J.~L.}\ \bibnamefont {Bl\'azquez-Salcedo}}, \bibinfo {author} {\bibfnamefont {F.~S.}\ \bibnamefont {Khoo}}, \ and\ \bibinfo {author} {\bibfnamefont {J.}~\bibnamefont {Kunz}},\ }\href {\doibase 10.1103/PhysRevD.96.064008} {\bibfield  {journal} {\bibinfo  {journal} {Phys. Rev. D}\ }\textbf {\bibinfo {volume} {96}},\ \bibinfo {pages} {064008} (\bibinfo {year} {2017})},\ \Eprint {http://arxiv.org/abs/1706.03262} {arXiv:1706.03262 [gr-qc]} \BibitemShut {NoStop}%
\bibitem [{\citenamefont {Konoplya}\ \emph {et~al.}(2019{\natexlab{b}})\citenamefont {Konoplya}, \citenamefont {Zinhailo},\ and\ \citenamefont {Stuchl\'\i{}k}}]{Konoplya:2019hml}%
  \BibitemOpen
  \bibfield  {author} {\bibinfo {author} {\bibfnamefont {R.~A.}\ \bibnamefont {Konoplya}}, \bibinfo {author} {\bibfnamefont {A.~F.}\ \bibnamefont {Zinhailo}}, \ and\ \bibinfo {author} {\bibfnamefont {Z.}~\bibnamefont {Stuchl\'\i{}k}},\ }\href {\doibase 10.1103/PhysRevD.99.124042} {\bibfield  {journal} {\bibinfo  {journal} {Phys. Rev. D}\ }\textbf {\bibinfo {volume} {99}},\ \bibinfo {pages} {124042} (\bibinfo {year} {2019}{\natexlab{b}})},\ \Eprint {http://arxiv.org/abs/1903.03483} {arXiv:1903.03483 [gr-qc]} \BibitemShut {NoStop}%
\bibitem [{\citenamefont {Bhagwat}\ \emph {et~al.}(2022)\citenamefont {Bhagwat}, \citenamefont {Pacilio}, \citenamefont {Barausse},\ and\ \citenamefont {Pani}}]{Bhagwat:2021kwv}%
  \BibitemOpen
  \bibfield  {author} {\bibinfo {author} {\bibfnamefont {S.}~\bibnamefont {Bhagwat}}, \bibinfo {author} {\bibfnamefont {C.}~\bibnamefont {Pacilio}}, \bibinfo {author} {\bibfnamefont {E.}~\bibnamefont {Barausse}}, \ and\ \bibinfo {author} {\bibfnamefont {P.}~\bibnamefont {Pani}},\ }\href {\doibase 10.1103/PhysRevD.105.124063} {\bibfield  {journal} {\bibinfo  {journal} {Phys. Rev. D}\ }\textbf {\bibinfo {volume} {105}},\ \bibinfo {pages} {124063} (\bibinfo {year} {2022})},\ \Eprint {http://arxiv.org/abs/2201.00023} {arXiv:2201.00023 [gr-qc]} \BibitemShut {NoStop}%
\bibitem [{\citenamefont {Berti}\ \emph {et~al.}(2006)\citenamefont {Berti}, \citenamefont {Cardoso},\ and\ \citenamefont {Will}}]{Berti:2005ys}%
  \BibitemOpen
  \bibfield  {author} {\bibinfo {author} {\bibfnamefont {E.}~\bibnamefont {Berti}}, \bibinfo {author} {\bibfnamefont {V.}~\bibnamefont {Cardoso}}, \ and\ \bibinfo {author} {\bibfnamefont {C.~M.}\ \bibnamefont {Will}},\ }\href {\doibase 10.1103/PhysRevD.73.064030} {\bibfield  {journal} {\bibinfo  {journal} {Phys. Rev. D}\ }\textbf {\bibinfo {volume} {73}},\ \bibinfo {pages} {064030} (\bibinfo {year} {2006})},\ \Eprint {http://arxiv.org/abs/gr-qc/0512160} {arXiv:gr-qc/0512160} \BibitemShut {NoStop}%
\bibitem [{\citenamefont {Berti}\ \emph {et~al.}(2016)\citenamefont {Berti}, \citenamefont {Sesana}, \citenamefont {Barausse}, \citenamefont {Cardoso},\ and\ \citenamefont {Belczynski}}]{Berti:2016lat}%
  \BibitemOpen
  \bibfield  {author} {\bibinfo {author} {\bibfnamefont {E.}~\bibnamefont {Berti}}, \bibinfo {author} {\bibfnamefont {A.}~\bibnamefont {Sesana}}, \bibinfo {author} {\bibfnamefont {E.}~\bibnamefont {Barausse}}, \bibinfo {author} {\bibfnamefont {V.}~\bibnamefont {Cardoso}}, \ and\ \bibinfo {author} {\bibfnamefont {K.}~\bibnamefont {Belczynski}},\ }\href {\doibase 10.1103/PhysRevLett.117.101102} {\bibfield  {journal} {\bibinfo  {journal} {Phys. Rev. Lett.}\ }\textbf {\bibinfo {volume} {117}},\ \bibinfo {pages} {101102} (\bibinfo {year} {2016})},\ \Eprint {http://arxiv.org/abs/1605.09286} {arXiv:1605.09286 [gr-qc]} \BibitemShut {NoStop}%
\bibitem [{Bar(2019)}]{Barack_2019}%
  \BibitemOpen
  \href {\doibase 10.1088/1361-6382/ab0587} {\ \textbf {\bibinfo {volume} {36}},\ \bibinfo {pages} {143001} (\bibinfo {year} {2019})}\BibitemShut {NoStop}%
\bibitem [{\citenamefont {Giesler}\ \emph {et~al.}(2019)\citenamefont {Giesler}, \citenamefont {Isi}, \citenamefont {Scheel},\ and\ \citenamefont {Teukolsky}}]{Giesler:2019uxc}%
  \BibitemOpen
  \bibfield  {author} {\bibinfo {author} {\bibfnamefont {M.}~\bibnamefont {Giesler}}, \bibinfo {author} {\bibfnamefont {M.}~\bibnamefont {Isi}}, \bibinfo {author} {\bibfnamefont {M.~A.}\ \bibnamefont {Scheel}}, \ and\ \bibinfo {author} {\bibfnamefont {S.}~\bibnamefont {Teukolsky}},\ }\href {\doibase 10.1103/PhysRevX.9.041060} {\bibfield  {journal} {\bibinfo  {journal} {Phys. Rev. X}\ }\textbf {\bibinfo {volume} {9}},\ \bibinfo {pages} {041060} (\bibinfo {year} {2019})},\ \Eprint {http://arxiv.org/abs/1903.08284} {arXiv:1903.08284 [gr-qc]} \BibitemShut {NoStop}%
\bibitem [{\citenamefont {Nee}\ \emph {et~al.}(2023)\citenamefont {Nee}, \citenamefont {V\"olkel},\ and\ \citenamefont {Pfeiffer}}]{Nee:2023osy}%
  \BibitemOpen
  \bibfield  {author} {\bibinfo {author} {\bibfnamefont {P.~J.}\ \bibnamefont {Nee}}, \bibinfo {author} {\bibfnamefont {S.~H.}\ \bibnamefont {V\"olkel}}, \ and\ \bibinfo {author} {\bibfnamefont {H.~P.}\ \bibnamefont {Pfeiffer}},\ }\href {\doibase 10.1103/PhysRevD.108.044032} {\bibfield  {journal} {\bibinfo  {journal} {Phys. Rev. D}\ }\textbf {\bibinfo {volume} {108}},\ \bibinfo {pages} {044032} (\bibinfo {year} {2023})},\ \Eprint {http://arxiv.org/abs/2302.06634} {arXiv:2302.06634 [gr-qc]} \BibitemShut {NoStop}%
\end{thebibliography}%

\begin{widetext}
\appendix

\section{Effective potentials}
\label{app:potentials}

In what follows, we report the explicit expressions of the radial potential for the scalar, electromagnetic, and gravitational perturbations pertaining to the class of nonlocal BH metrics \eqref{eq:BH_sol}.
\begin{align}
    V^{(0)}&= \frac{(r-2) \left(2+r\ell+r\ell^2\right)}{r^4}+\alpha\bigg\{\frac{3 \left[k (12-19 r)-18 (r-1)+2 k^2 (5 r-3)\right]}{r^{k+3}(r-3)^3}+\frac{3(36-54 r+30 r^2-8 r^3+r^4)}{3^kr^4(r-3)^3}\notag \\
    &+\frac{40-6 r+k (63-14 r+r^2)-k^2 (37-10 r+r^2)}{2r^{k+1}(r-3)^3}-\frac{ \ell (\ell+1)}{r^{k+2}}\bigg\}\,, \\
    \label{eq:potential_scalar}
    V^{(1)}&= \frac{\ell (\ell+1) (r-2)}{r^3}-\alpha\left[\frac{\ell (\ell+1)}{ r^{k+2}}\right], \\
    V^{(2)}&= \frac{(r-2) \left(\ell^2 r+\ell r-6\right)}{r^4}+\alpha\bigg\{\frac{2 \left(6-r\ell-r\ell^2\right)}{r^{k-1}}+\frac{r-2}{(3r)^k(r-3)^3}\Big[324 r^k-432 r^{k+1}+192 r^{k+2}-30 r^{k+3} \notag \\
    & +3^k k (k+3) r^4-2\times 3^k \left(4 k^2+10 k-9\right) r^3+3^{k+1} \left(7 k^2+15 k-28\right) r^2-2\times 3^{k+2} \left(k^2+2 k-6\right) r\Big]\bigg\}.
    \label{eq:potential_gravitational}
\end{align}

It is worth to emphasize that none of the above potentials exhibits a singularity at $r=3$, as demonstrated by the following finite limits:
\begin{align}
    \lim_{r\to3} V^{(0)}&=\frac{1}{2\times3^{5 + k}} \Big\{4 \times 3^{1 + k} + 
18\ell (\ell + 1) \left( 3^k - 3\alpha \right) + \alpha \big[k \left( 11 + (k-6)k \right)-24 \big] \Big\}\,, \label{eq:limit-scalar}\\
\lim_{r\to3} V^{(1)}&=\frac{\ell (\ell + 1) \left(3^k - 3\alpha \right)}{3^{3+k}}\,, \label{eq:limit-electromagnetic}\\
\lim_{r\to3} V^{(2)}&= \frac{1}{486} \left\{18 \left(\ell^2+\ell-2\right)-\frac{\alpha}{ 3^k} \left[k^3+17 k+54 \ell (\ell+1)-72\right]\right\}\,. \label{eq:limit-gravitational}
\end{align}

\begin{figure*}[b]
    \centering
    \includegraphics[width=0.32\linewidth]{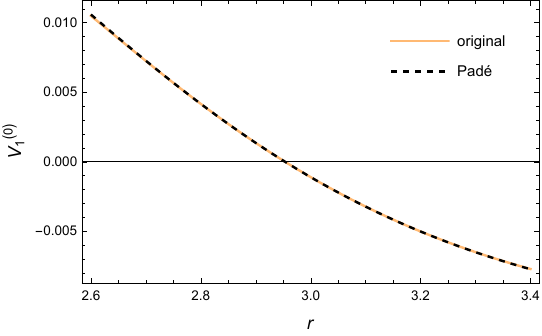}\quad 
    \includegraphics[width=0.32\linewidth]{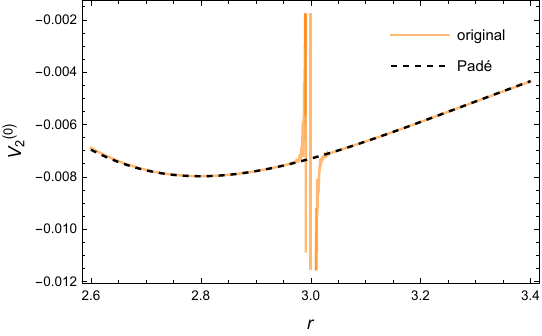}\quad
    \includegraphics[width=0.32\linewidth]{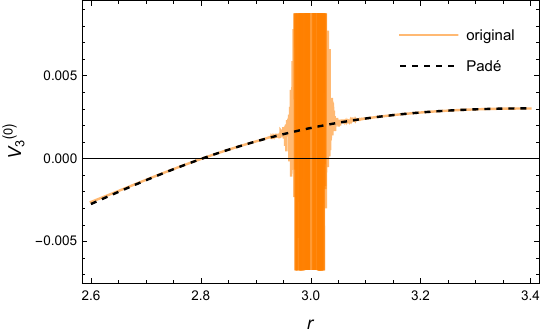}
    \\ \vspace{0.5cm}
    \includegraphics[width=0.32\linewidth]{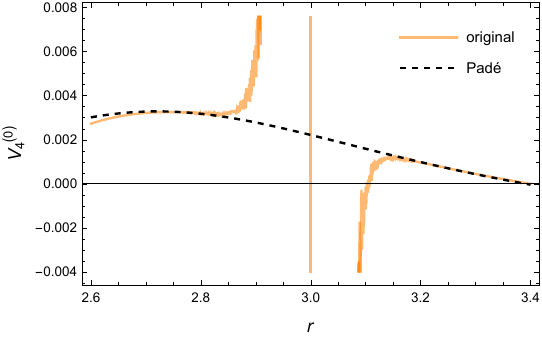}\quad
    \includegraphics[width=0.32\linewidth]{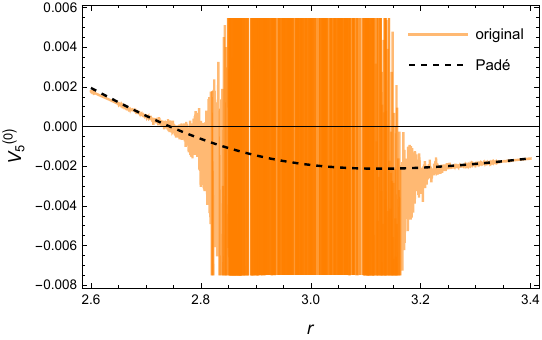}\quad
    \includegraphics[width=0.32\linewidth]{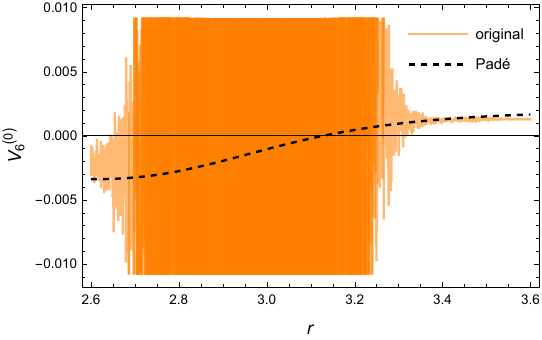}
    \caption{Pad\'e approximants of the derivatives of the effective potential for scalar perturbations for $n=1$, $\ell=1$, $\alpha=0.1$ and $k=2$.}
    \label{fig:Pade_scalar}
\end{figure*}

\begin{figure*}
    \centering
    \includegraphics[width=0.32\linewidth]{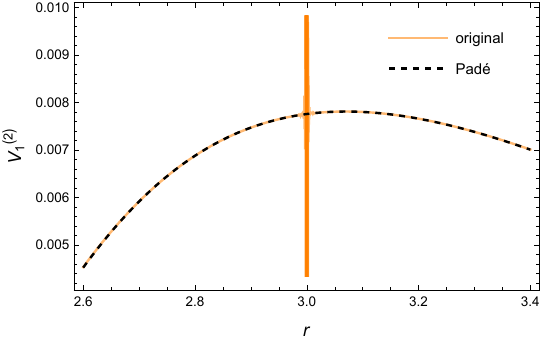}\quad 
    \includegraphics[width=0.32\linewidth]{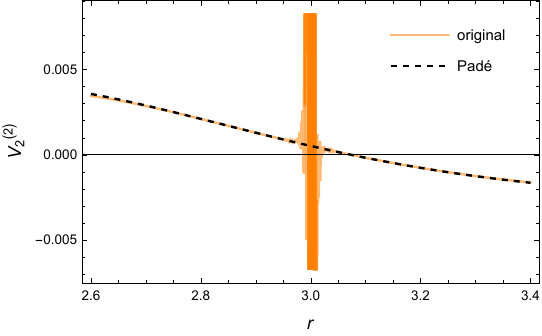}\quad
    \includegraphics[width=0.32\linewidth]{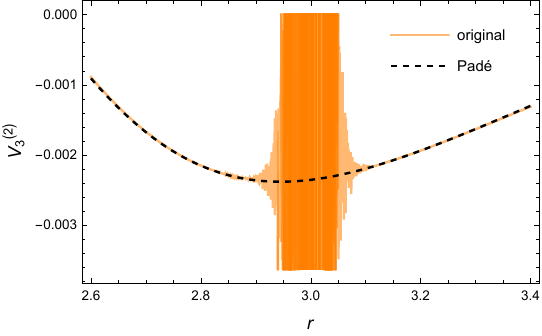} 
     \\ \vspace{0.5cm}
    \includegraphics[width=0.32\linewidth]{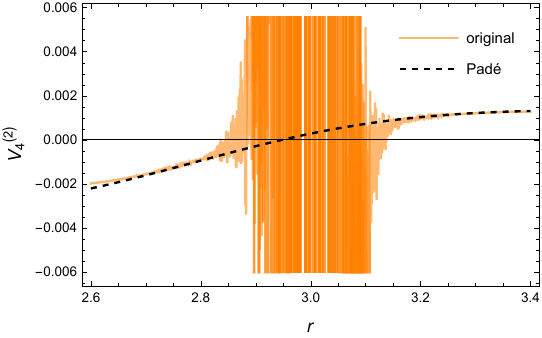}\quad
    \includegraphics[width=0.32\linewidth]{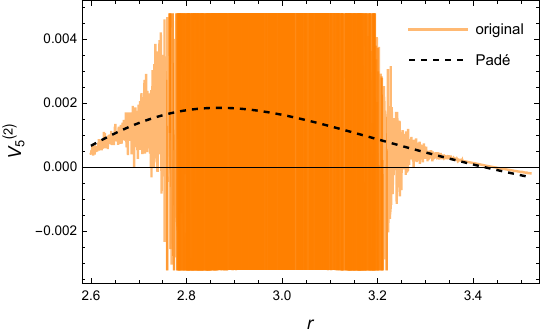}\quad
    \includegraphics[width=0.32\linewidth]{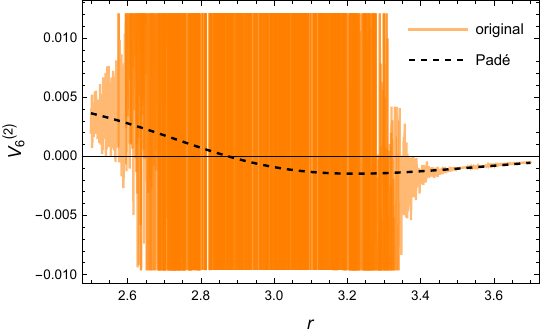}
    \caption{Pad\'e approximants of the derivatives of the effective potential for gravitational perturbations for $n=1$, $\ell=1$, $\alpha=0.1$ and $k=2$. }
    \label{fig:Pade_gravitational}
\end{figure*}

\section{Pad\'e approximations}
\label{app:Pade}

To illustrate the accuracy of the Pad\'e method, we display in Figs.~\ref{fig:Pade_scalar} and \ref{fig:Pade_gravitational} the Pad\'e approximations of the derivatives of the effective potential for scalar and gravitational BH perturbations around $r=3$, respectively. 
We observe that the Pad\'e method provides an effective analytical treatment of the numerical instabilities arising near the peak of the effective potential's derivatives, while ensuring a smooth and accurate match with the original expressions away from this region.
Although the precise degree of approximation is difficult to assess in this context, the Pad\'e technique is known for its optimal convergence and stability properties \cite{Capozziello:2020ctn}, which support the reliability of our final results.   

\end{widetext}

\end{document}